\documentclass[letterpaper,11pt]{article}
\pdfoutput=1 % if your are submitting a pdflatex (i.e. if you have images in pdf, png or jpg format)

\usepackage{jheppub} % JHEP

\usepackage{epsf,epsfig,subfigure,graphicx,amsmath,amssymb,hyperref,cancel}
\usepackage{subfigure}

\newcommand{\be}{\begin{equation}}
\newcommand{\ee}{\end{equation}}
\newcommand{\bea}{\begin{eqnarray}}
\newcommand{\eea}{\end{eqnarray}}

\def\beq#1\eeq{\begin{align}#1\end{align}}
\def\beqnn#1\eeq{\begin{align*}#1\end{align*}}
\newcommand{\nn}{\nonumber}
\newcommand{\rpv}{$R$-parity violation}
\newcommand{\rpa}{$R$-parity}
\newcommand{\rpcing}{$R$-parity conserving\ }
\newcommand{\rpving}{$R$-parity violating\ }

\def \mtr { m_{3/2}}

\def \nta { n_{\tilde a}}
\def \th {_{3/2}}
\def \l {\lambda''}
\newcommand{\fru}[2]{\left( \frac{#1}{\text{#2}}\right)}
\newcommand{\frd}[2]{\left( \frac{\text#1}{{#2}}\right)}
\newcommand{\frud}[2]{\left( \frac{#1}{{#2}}\right)}

\def \mta { m_{\tilde a}}

\def \mtq { m_{\tilde q}}
\def \mtg { m_{\tilde g}}

\def \tev {\text{ TeV}}
\def \gev {\text{ GeV}}
\def \mev {\text{ MeV}}

\newcommand{\OO}{\mathcal{O}}
\newcommand{\LL}{\mathcal{L}}

\def\nanow{Nanopoulos-Weinberg}
\def\bnving{baryon-number-violating }

\begin{document}

%\begin{frontmatter}
\preprint{RUNHETC-2014-22}

\title{Axino LSP Baryogenesis and Dark Matter}
\author{Angelo Monteux,}
\author{Chang Sub Shin}
\affiliation{New High Energy Theory Center, Department of Physics and Astronomy,
\\Rutgers University, Piscataway, NJ 08854, USA
}
\emailAdd{amonteux@physics.rutgers.edu  }
\emailAdd{changsub@physics.rutgers.edu}

%\begin{abstract}
\abstract{
We discuss a new mechanism for baryogenesis, in which the baryon asymmetry is generated by the lightest particle in another sector, for example the supersymmetric particle (LSP), decaying to quarks via baryonic-number-violating interactions. As a specific example, we use a supersymmetric axion model with an axino LSP and baryonic $R$-parity violation. This  scenario predicts large  $R$-parity violation for the stop, and an upper limit on the squark masses between {15 and 130 TeV}, for different choices of the Peccei-Quinn scale and the soft $X_t$ terms. We discuss the implications for the nature of dark matter in light of the axino baryogenesis mechanism, and find that both the axion and a metastable gravitino can provide the correct dark matter density. In the axion dark matter scenario, the initial misalignment angle is restricted to be ${\cal O}(1)$.  On the other hand, the reheating temperature  is linked to the PQ scale and should be higher than $10^4-10^5$ GeV in the gravitino dark matter scenario.
}

%Thermal production of gravitino is dominant compared to the non-thermal production by axino decays. 
% and find that both the axion and an metastable gravitino can easily provide the correct relic density. %In both cases, the prospects for direct detection experiments are dim. %The allowed gravitino masses and Peccei-Quinn scale are
%\end{abstract}

%\begin{keyword}
%\texttt{elsarticle.cls}\sep \LaTeX\sep Elsevier \sep template
%\MSC[2010] 00-01\sep  99-00
%\end{keyword}

%\end{frontmatter}

\maketitle

%%%%%%%%%%%%%%%%%%%%%%%%%%%%%%%%%%
\section{Introduction}

Two fundamental observations about our universe, namely the presence of dark matter and the prevalence of matter over antimatter, %have been investigated 
can individually find an explanation in theories that go beyond the Standard Model. However, it is harder to explain both in a unified framework.
%\footnote{But possible, as in the class of models inspired by \cite{Cui:2012jh}.}
A simple argument based on symmetries can justify this: a stable dark matter candidate would be protected by a symmetry, while baryogenesis requires the violation of another symmetry, namely baryon number (or lepton number in high-energy leptogenesis). At the same time, flavor physics is sensitive to baryon-number violation, and proton decay is mediated by the simultaneous breaking of baryon and lepton number. 

Although the symmetries involved in the two mechanisms need not to be related, they are in supersymmetric theories, where \rpa\ forbids renormalizable baryon-number-violating operators and at the same time provides a stable dark matter candidate, the Lightest Supersymmetric Particle (LSP). The LSP is also the source of the missing energy signature of supersymmetric events at colliders. 
Given the stringent LHC limits for light \rpcing supersymmetry (SUSY) and the null results in dark matter direct detection experiments, it is appealing to investigate phenomenological consequences of \rpv\, (RPV). 
%Besides collider studies, its cosmological implication would be very interesting because one can find some clues to relate  generation mechanism of baryons and that of dark matters.  
Besides collider studies, its cosmological implication is very interesting because one can relate the baryogenesis mechanism and the nature of  dark matter.  

Sizable lepton and proton number violation are not consistent with bounds on the proton lifetime, thus we will focus on models allowing only for baryonic \rpv. In addition to the Minimal Supersymmetric Standard Model (MSSM) superpotential, we include the operator
\beq\label{udd}
\frac12 {\lambda''_{ijk}}U_i^c D_j^c D_k^c
\eeq 
where the RPV couplings $\l_{ijk}$ are antisymmetric under the exchange $j\leftrightarrow k$ and the sum over color indices is understood. %Because this operator violates baryon number, mechanisms for baryogenesis are easily constructed with the out-of-equilibrium decay of superpartners \cite{Dimopoulos:1987rk,Cline:1990bw,Mollerach:1991mu}. At the same time,  
Neutralinos decay away in the early universe.   
Without an additional symmetry to warrant dark matter stability, 
a natural candidate for  dark matter should be super-weakly interacting: for example, the gravitino, which can be metastable and  have a lifetime longer than the age of the universe, and the axion, which was originally introduced \cite{Peccei:1977hh} to explain the absence of CP-violation in strong interactions (the strong CP problem). 
Introducing the axion supermultiplet gives a further implication that relates dark matter and baryogenesis; there is a fermionic superpartner, the axino, whose out-of equilibrium decay through the operator \eqref{udd} can generate a baryon asymmetry. 

%Previous works linking dark mater and baryon asymmetry include \cite{Kaplan:2009ag,Cui:2012jh,Arcadi:2013jza}. 
The idea of using the operator \eqref{udd} for baryogenesis is not new. Refs. \cite{Dimopoulos:1987rk,Cline:1990bw,Mollerach:1991mu} discussed a baryon asymmetry generated at low temperatures (down to the MeV scale) from late decays of inflaton, gravitinos and axinos into superpartners, respectively\footnote{In another class of models that gives an asymmetry through the operator \eqref{udd} \cite{Cui:2012jh,Cui:2013bta,Arcadi:2013jza,RompineveSorbello:2013xwa}, 
 the decaying particles are thermally produced from freeze-out at a temperature well below the mass of the particles.}.
It is important that baryon asymmetry is generated at a temperature well below the superpartner mass scale so that it is not washed out by baryon-number-violating processes in the thermal bath \cite{Dreiner:1992vm}. For large RPV couplings,  baryogenesis should happen at a temperature lower than about $\mtq/20$. Otherwise, if the asymmetry is generated at or above the superpartner scale, all the \rpving couplings have to be smaller than $\OO(10^{-7})$ for baryons to survive. 

We shortly review these baryogenesis mechanisms, which all involve \rpving $A$-terms. In  \cite{Dimopoulos:1987rk},  out-of-equilibrium decays of the inflaton generate a non-thermal population of squarks, which later decay to a quark and a gluino, provided that  $\mtq>\mtg+m_q$. Interference between the tree level and two-loop diagrams  gives a baryon asymmetry. In  \cite{Cline:1990bw}, the decay of a gravitino to a quark and a squark (or to a gluino and a gluon, with the subsequent gluino decay to a quark and a squark)  was used to generate the asymmetry (given the hierarchies $\mtr>\mtq+m_q$ or $\mtr> \mtg> \mtq+m_q$, respectively), through a one-loop diagram involving $A$-terms. In  \cite{Mollerach:1991mu}, the same diagrams were used to discuss the asymmetry generated by the decay of the axino (or saxion) to a gluino and finally to a quark and a squark.

For the cases described above, the parent particle decays to some on-shell superpartner via \rpcing interactions, and interference with a \rpving loop diagram generates the baryon asymmetry. A specific hierarchy is required in each case: the particle decaying out of equilibrium is always heavier than the  superpartners that are on-shell in the interfering diagram. This automatically excludes the case of an LSP decay (in the following the LSP is defined as in $R$-parity conserving SUSY, and it is stable only when the RPV interactions are turned off). In fact, this reflects a more general statement: Nanopoulos and Weinberg proved in Ref.  \cite{Nanopoulos:1979gx} that an LSP defined in such a way cannot give a baryon asymmetry at first order in the \bnving interactions.

In this work, we will investigate how to generate both a baryon asymmetry at low energies and a correct dark matter density.  We focus on models with baryonic RPV and show a new baryogenesis mechanism in which a late-decaying axino LSP gives rise to the observed baryon asymmetry (at second order in the \rpving interactions). Baryogenesis takes place below the weak scale and before Big Bang Nucleosynthesis. Both gravitino and axion dark matter are discussed, with the nature of dark matter constraining the parameter space for baryogenesis and vice versa.

This paper is organized as follows: we recall the \nanow \ theorem in section \ref{sec:NWtheorem}, and discuss the implications for baryogenesis through an LSP decay. In section \ref{sec:baryogenesis} we derive a new mechanism that generates substantial baryon asymmetry through late decays of an axino LSP. In section \ref{sec:dm}, we investigate the possibilities for dark matter candidates and discuss the range of parameter space in which both a baryon asymmetry and a correct dark matter density exist. We discuss the collider bounds on our model and conclude in section \ref{sec:conclusions}.

%\section{\rpv\  and Baryogenesis}\label{sec:NWtheorem}
%\subsection{The Nanopolous-Weinberg theorem and LSP baryogenesis}
\section{The Nanopoulos-Weinberg theorem and LSP baryogenesis}\label{sec:NWtheorem}

In Ref.  \cite{Nanopoulos:1979gx} it was shown that, given a particle $X$ which is stable in the limit of no \bnving interactions, the decay rate of  $X$ into all final states with a given baryon number $B$ equals the decay rate of its antiparticle $\overline X$ into all states with baryon number $-B$,  at first order in the \bnving interactions. Because the net number of baryons is proportional to the difference of the two decay rates, no baryon asymmetry can be generated by an LSP decay at first order. 

The \nanow \ theorem was further investigated in Ref.  \cite{Adhikari:2001yr}, where it was generally argued that a difference in the two decay rates  exists only if the on-shell intermediate particles and the final particles have a different baryon number. In other words, the process to the right of the ``cut'' must violate baryon number. This results holds at all orders in the \bnving interactions. For an LSP decay, \bnving operators must also appear on the left of the ``cut'' to have on-shell intermediate particles. If the parent particle is not the LSP (it has baryon-number-conserving decay channels), it is possible to have an asymmetry at first order, with a baryon-number-conserving interaction to the left of the ``cut''. 
The models of  \cite{Dimopoulos:1987rk,Cline:1990bw,Mollerach:1991mu}, where baryon asymmetry is generated by two-body decays, and of \cite{Cheung:2013hza}, where it comes from three-body decays,
%via higher dimension operators,
 fall in this last category, as they used decays of heavier particles, and were indeed able to get an asymmetry at first order in the RPV couplings.
% all the wrong LSP papers get the silence treatment
%\footnote{As does Ref. \cite{CC}, which treated the (baryon-number-violating) three-body decay of a field $X_I$ interacting via higher dimension operators, %also falls in this category, %as it allowed for the decay of $X_I$ to a lighter field $ X_J$.
%}

For the LSP case, it is natural to consider a low-energy effective theory containing the LSP $\chi$, a  Majorana fermion with mass $m_\chi$, and the Standard Model fields, where all the heavier degrees of freedom have been integrated out. Baryon number violation is present in non-renormalizable operators which are suppressed by the heavier particles masses (in RPV SUSY, the squarks). The effective operators involving the quarks $q$ and the LSP will be of the schematic form
\beq
\LL_{|\Delta B|=1}\sim \frac{c_1}{\Lambda^2}(\chi q)(q q)+h.c.,\qquad \LL_{|\Delta B|=2}\sim \frac{c_2}{\Lambda^5}(qq)(qq)(qq)+h.c.\,,
\eeq
where the gauge indices are contracted to form gauge singlets and $\Lambda$ is the mass scale of the heavier particles. Then the decay rate of $\chi$ becomes 
\bea
\Gamma_{\chi\to qqq} \sim \frac{1}{ (8\pi)^3}\frac{|c_1|^2m_\chi^5}{\Lambda^4}.
\eea
\begin{figure}[t]
\begin{center}
\includegraphics{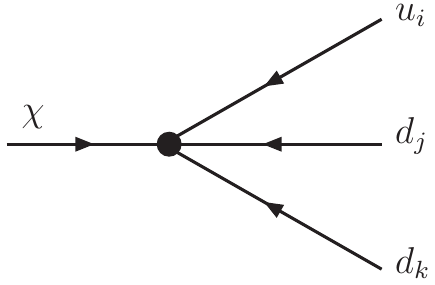}\hspace{2cm}
\includegraphics{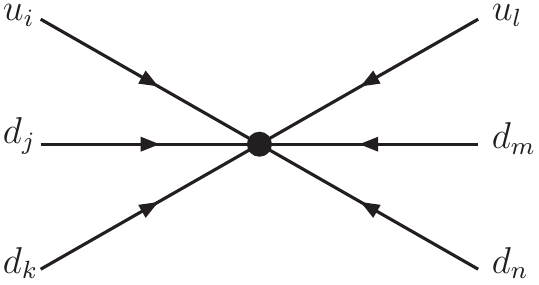}
\caption{Non-renormalizable $\Delta B=1$ and $\Delta B=2$ effective operators in the low-energy theory. The two-loop diagram with on-shell intermediate quarks is obtained by contracting the operators together, with  one $m_\chi$ insertion.}
\label{fig:deltaB_EFT}
\end{center}
\end{figure}
An asymmetry between $\Gamma_{\chi\to qqq}$ and $\Gamma_{\chi\to \bar q\bar q\bar q}$ will come from the interference between the tree level $\Delta B=1$ decay and the two-loop decay diagram  { obtained joining together the two operators in Fig.  \ref{fig:deltaB_EFT}:}%in both $\LL_{|\Delta B|=1}$ and $\LL_{|\Delta B|=2}$:
\footnote{
{Including the Majorana mass term, the theory has three complex parameters and two independent fields ($\chi$ and $qqq$), resulting in one phase that cannot be removed by field redefinition of the $q$'s and $\chi$. We can check that this is the phase appearing in $\epsilon$ by considering the invariant combination of operators under the field redefinition, $(c_1^* \bar\chi \bar q\bar q\bar q)^2 (c_2 qqqqqq) (m_\chi \chi\chi)$.}}
\bea
\epsilon\equiv \frac{\Gamma_{\chi\to qqq} - \Gamma_{\chi\to \bar q\bar q\bar q}}{\Gamma_{\chi\to qqq} + \Gamma_{\chi\to \bar q\bar q\bar q}}
\sim \frac{1}{(8\pi)^3 }\frac{{\rm Im}{ [c_1^{*2} c_2 m_\chi]\, m_\chi^2}}{|c_1|^2 \Lambda^3}.
\eea
It is worth to note that the asymmetry is generated not only at second order in the \bnving interactions (as expected by the \nanow\ theorem), but also at two-loops: a 1-loop diagram would require a dimension six, $\Delta B=2$ operator in the effective theory, which is not allowed. 
Finally, the present baryon asymmetry depends on the LSP abundance at decay time,
\beq\label{baryon_asymmetry}
Y_{\Delta B}\equiv \frac{n_B}s=\epsilon \frud{n_\chi}s _{t= 1/\Gamma_\chi}.
\eeq
This can be compared to its experimental value \cite{Ade:2013zuv}, $(Y_{\Delta B})_{obs}=(0.80\pm0.018) \times 10^{-10}$.

In the next section, we will see the axino LSP is a good example to realize this mechanism.

\section{Baryogenesis from axino decays}\label{sec:baryogenesis}

In this section we present a concrete example of the LSP baryogenesis scenario just outlined. First, the particle has to decay out of thermal equilibrium. A typical candidate would be an LSP gravitino, % which could decay to baryons when heavier than a few GeV. Gravitinos are
 produced in the reheating  epoch \cite{Moroi:1993mb,Bolz:2000fu}: without \rpa, gravitinos decay to three quarks via the RPV operator of eq. \eqref{udd}. Compared to a non-LSP gravitino, this decay is much slower, as it is suppressed by the intermediate squark mass, by the RPV coupling $\l$ and by the three-body kinematic factor. In order for the decay products not to interfere with Big Bang Nucleosynthesis (BBN), the gravitino LSP should be extremely heavy, $\mtr\gtrsim 10^3 \tev$, with the other superpartners  being even heavier \cite{Moreau:2001sr}. Even so, all superpartners would have decayed in the early universe and there would be no dark matter candidate. %One could consider axion dark matter, but in that case an axino would also be introduced. As we will see, the axino can generate a baryon asymmetry on its own, in a more natural region of the parameter space.

Thus, we examine a supersymmetric QCD axion model in which 
the axion ($a$), the pseudo-Goldstone boson associated with the spontaneous breaking of an anomalous Peccei-Quinn (PQ) $U(1)$ symmetry at a scale $v_{PQ}$, solves the strong CP problem of QCD \cite{Kim:2008hd} and is a good dark matter candidate. Assuming that the axino ($\tilde a$), the  fermionic superpartner of the axion, is  the LSP, a  baryon  asymmetry  can be generated  by  its  decays in a more natural region of the parameter space than in the gravitino LSP case.  
The large value of $v_{PQ}$ ensures that axinos are out-of-equilibrium when they decay, and the lifetime is much longer than the period in which RPV interactions are in thermal equilibrium (thus, the asymmetry is not washed out). 
 The chiral axion superfield can be written as  \bea
A= \frac{1}{\sqrt{2}}(s + i a) + \sqrt{2}\theta\tilde a  + \theta^2 F^A,
\eea
where the saxion, the scalar superpartner of the axion, is denoted as $s$. We consider a following superpotential  terms for $A$ to give interactions with the MSSM particles:
\bea
\Delta W = e^{q_H A/v_{PQ}} \mu H_u H_d + e^{q_\Phi A/v_{PQ}} M_{\Phi} \Phi \Phi^c ,
\eea
where $H_u,\,H_d$ are the MSSM Higgs doublets, and 
 $\Phi,\,\Phi^c$ are SM charged matter fields with $M_\Phi ={\cal O}(v_{PQ})$.
   The $U(1)_{PQ}$ symmetry is realized as  $A\to A+ i\theta v_{PQ}$,  $H_uH_d\to e^{-i q_H \theta}H_uH_d$,  and  $\Phi\Phi^c\to e^{- iq_\Phi \theta}\Phi\Phi^c$.
  This is a hybrid of the  DFSZ \cite{Dine:1981rt,Zhitnitsky:1980tq} and KSVZ \cite{Kim:1979if,Shifman:1979if} axion models, 
  in which the axino  decay is dominated by the first term as in the DFSZ case while 
  at high temperature
its thermal production is mostly given by that of the KSVZ model \cite{Bae:2014efa}. 

\begin{figure}[t]
\centering
\subfigure[]{\includegraphics{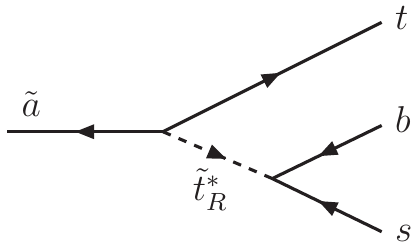}}\qquad\qquad
\subfigure[]{\includegraphics{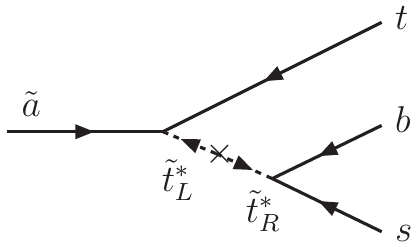}}\\
\subfigure[]{\includegraphics{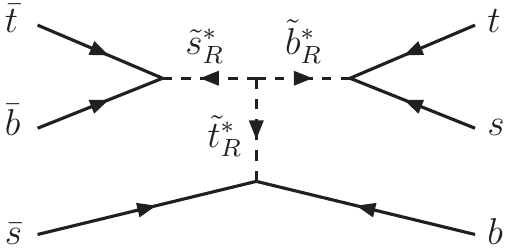}}\qquad\quad
\subfigure[]{\includegraphics{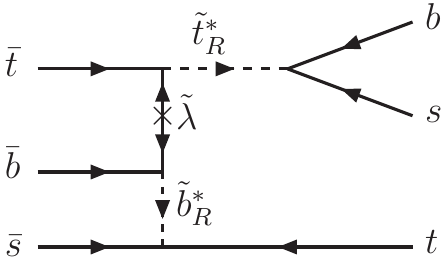}}
\caption{Diagrams leading to $\Delta B=1$ and $\Delta B=2$ operators as in Fig. \ref{fig:deltaB_EFT}, but in the full theory. (a) Tree-level $\Delta B=1$ axino decay. (b) Same decay, involving left-right squark mixing ($X_t$). (c) $\Delta B=2$  process involving the RPV $A$-term. (d)  $\Delta B=2$ process mediated by a gaugino in the $t$-channel; here, there are two  similar diagrams with the exchange exchange $\tilde t_R \to \tilde b_R,\tilde s_R$, plus three analogous diagrams with the gaugino in the $s$-channel.
}
\label{fig:axino_tree}
\end{figure}

Because all the sparticles are heavier, we can consider a low-energy effective theory with SM quarks supplemented by the axino, a Majorana particle with a mass $m_{\tilde a}$. 
Non-renormalizable interactions for the quarks remain after integrating out the squarks in the diagrams of Fig. \ref{fig:axino_tree}. 
The following effective axino interactions are given by the  mixing of the axino with the higgsino, after integrating out the squarks:
\beq
{\cal L }_{|\Delta B|=1} =& \sum_\alpha \frac{ \kappa_\alpha m_{u_\alpha}}{v_{PQ}}
\frac{\l_{\alpha\beta\gamma}}{m_{\tilde u_{R\alpha}}^2} (\bar {\tilde a} \bar u_\alpha)(d^c_\beta d^c_\gamma)  + h.c.  \nonumber \\\label{Ldb1}
&+ \sum_\alpha\frac{\kappa_\alpha m_{u_\alpha} }{v_{PQ}} 
\frac{\l_{\alpha\beta\gamma}}{m_{\tilde u_{R\alpha}}^2}
\frac{ m_{u_\alpha} X_{u_\alpha}}{m_{\tilde u_{L\alpha}}^2}
(\tilde a u_\alpha^c)(d^c_\beta d^c_\gamma) + h.c.\,,
\eeq
where $q_\alpha$ (respectively, $q_\alpha^c$) are the left-handed (right-handed) quarks.
The holomorphic term in the second line comes from the left-right squark mixing, $X_{u_\alpha} \equiv A_{u_\alpha}+ \mu \cot\beta$. $\kappa_\alpha$ is an $\OO(1)$ coefficient given by the charges of the SM fields under the PQ symmetry.\footnote{
In pure KSVZ models  \cite{Kim:1979if,Shifman:1979if}, the SM fields are neutral under the PQ symmetry and the same operator would arise at 1-loop after integrating out the gluino; it would be further suppressed with respect to the DFSZ case.}

The six-fermion holomorphic $\Delta B=2$ Lagrangian is obtained from the  soft SUSY breaking $A$-terms, $\Delta {\cal L}_{\rm soft} = \l_{ijk} A''_{ijk} \tilde u_{Ri} \tilde d_{Rj} \tilde d_{Rk} + h.c.$, after integrating out the right-handed squarks,
\beq
{\cal L}^A_{|\Delta B|= 2} =- \sum_{ijk,\alpha\beta\gamma,\delta\epsilon\zeta}\frac{\lambda_{ijk}^*A''^*_{ijk} \lambda_{i\epsilon\zeta} \lambda_{\delta j \gamma} 
\lambda_{\alpha \beta k }}{m_{\tilde u_{Ri}}^2 m_{\tilde d_{Rj}}^2 m_{\tilde d_{Rk}}^2} 
( d^c_\epsilon d^c_\zeta) (u^c_\delta d^c_\gamma)( u^c_\alpha d^c_\beta ) + h.c.,
\eeq
{ and also from squark-quark-gaugino interactions, $\Delta {\cal L}_{\rm susy} = 
\sqrt{2} g_A( \tilde u_{Ri}^*T^A u_{i}^c + \tilde d_{Ri}^* T^Ad_{i}^c)\tilde \lambda^A  + h.c.$, after integrating out the gauginos:
\beq \label{DeltaB2_gaugino}
{\cal L}^{\tilde \lambda}_{|\Delta B|=2} =&  - 
 \sum_{A,\alpha\beta\gamma,\delta\epsilon\zeta} \frac{ g_A^2 m_{\tilde \lambda^A}^*\lambda_{\alpha\beta\gamma} \lambda_{\delta\epsilon\zeta}}{4\,m_{\tilde \lambda^A}^2\, m_{\tilde u_{R\alpha}}^2 m_{\tilde u_{R\delta}}^2 }
(d^c_\beta d^c_\gamma)((T^A u^c_\alpha) (T^A u^c_{\delta}))(d^c_\epsilon d^c_\zeta)+ h.c.
\nonumber\\
&+ 4\left(\tilde u_{R\alpha}\to \tilde d_{R\gamma}, 
u^c_\alpha \leftrightarrow d^c_\gamma, \tilde u_{R\delta}\leftrightarrow  \tilde d_{R\zeta}, u^c_\delta
\leftrightarrow d^c_\zeta\right) \nonumber\\
&+ 2\left(\tilde u_{R\alpha}\leftrightarrow \tilde d_{R\gamma}, u^c_\alpha\leftrightarrow d^c_\gamma\right),
\eeq
where {$A$ indicates the gauge group index of the Standard Model, $\tilde\lambda^A$ 
is the corresponding gaugino with Majorana mass $m_{\tilde \lambda_A}^2$.  In the denominator} $m_{\tilde \lambda^A}^2$ is used to imply $|m_{\tilde \lambda^A}|^2$. 
If the gaugino masses are all of the same order, the gluino contribution is most important. 
If they follow the GUT relation as 
$m_{\tilde B}^2:\, m_{\tilde W}:\, m_{\tilde g}^2 \simeq 1: 4: 36$, the bino contribution would be also important. 
More detailed expressions are presented in the Appendix. 
Exchange of neutral Higgsinos also generates $\Delta B=2$ operator. However  
this contribution is proportional to
$y_uy_d\sim m_u m_d/v^2$, so  it is suppressed.
For the sake of simplicity, we neglect it and decouple the Higgsinos in our discussion, so that the bino Majorana mass term is nearly the mass eigenvalue of the neutralino.}

Because in Eq. \eqref{Ldb1} the effective axino coupling to quarks is proportional to the up-type quark masses, the dominant one involves the top quark. The relevant \rpving couplings are $\l_{312},\l_{313},\l_{323}$. For simplicity, we take the assumption that $\l_{323}$ is the only dominant coupling.\footnote{Note that we do not require all the other couplings to be zero, just to be small. This is justified in models where flavor symmetries determine a hierarchical structure of the \rpving couplings  \cite{Monteux:2013mna,Nikolidakis:2007fc,Csaki:2011ge}. The other couplings would be suppressed by spurions of the flavor symmetry.
}

%Assuming $\mta\gg m_t$, 
In the limit of massless final states,
the tree-level decay rate of the axino LSP %(Fig. \ref{fig:axino_tree}) 
is
\beq\label{eq:Gaxino}
\Gamma_{\tilde a\to tbs} &\simeq \frac{\kappa_3^2 |\lambda_{323}|^2}{{ 512}\pi^3}\frac{m_{\tilde a}^5 m_t^2}{v_{PQ}^2 m_{\tilde t_R}^4}\left[1+\frud{X_tm_t}{m_{\tilde t_L}^2}^2
\right]\,.
\eeq
Thus the  total decay rate is $\Gamma_{\tilde a} =\Gamma_{\tilde a\to t bs}+ \Gamma_{\tilde a\to \bar t \bar b\bar s} \approx 2 
\Gamma_{\tilde a\to t bs}$. Corrections to this result are proportional to $m_t^2/\mta^2$ and 
$m_{\tilde a}^2/m_{\tilde t_R}^2$  and are shown in the Appendix.

{
In order to obtain the baryon asymmetry as in Eq. \eqref{baryon_asymmetry},
we consider the axino thermal history  and evaluate its abundance in Sec.~(\ref{subsec:Axino_cosmology}). %evaluate the axino abundance first  in Sec.~(\ref{subsec:Axino_cosmology}), 
%considering its thermal history. 
The asymmetry parameter $\epsilon$ is presented in Sec.~(\ref{subsec:Axino_baryogenesis}), 
and we discuss its implications for the sparticle spectrum in the same section.}\\
%{\bf or}
%{Before computing the asymmetry parameter $\epsilon$, we obtain the axino decay temperature and abundance in the next section. }  

\subsection{Axino cosmology}\label{subsec:Axino_cosmology}

Even if the interactions between the axinos and  the MSSM particles are quite suppressed,  axinos can 
be generated from the thermal bath in the early Universe. 
For  a reheating temperature ($T_R$) much higher than the MSSM sparticle masses and much lower than the PQ breaking scale ($\mu\ll T_R \ll M_\Phi\sim v_{PQ}$),  axinos are mainly produced by scattering processes mediated by gluinos 
\cite{Covi:2001nw,Brandenburg:2004du,Strumia:2010aa,Chun:2011zd,Bae:2011jb,Bae:2011iw}. The axino yield from 
this thermal production is %in the same way  gravitinos are \cite{Strumia:2010aa}. 
\beq\label{axinoyield}
\left(\frac{\nta}s\right)_{TP}= 
{\rm min}\left[0.001 g_3^6 \left(\frac{200}{g_*(T_R)}\right)^{3/2} 
\left(\frac{T_R}{10^{10}\, {\rm GeV}}\right)\left(\frac{10^{12}\, {\rm GeV}}{v_{PQ}}\right)^2,
 0.002\left(\frac{200}{g_*(T_{\rm dec.})}\right)\right]
\,,
\eeq
where  ${\nta}$ is the axino number density, $s$ the entropy density of the Universe given by $(2\pi^2/45)g_{*s}T^3$ and 
$g_*(T)\simeq g_{*s}(T)$ is the number of relativistic degrees of freedom at the temperature $T$.
 If the reheating temperature is high enough ($T_R > T_{\rm dec.}$), 
 the scattering processes can be in chemical equilibrium. 
 In such a case, the axino number density  is $n_{\tilde a} = (3\zeta(3)/2\pi^2) T^3$ before it decouples, 
 and the corresponding yield after decoupling ($T < T_{\rm dec.}$) is given by the second term of the RHS of Eq. \eqref{axinoyield}.
%As for the gravitino case, the axino yield is dominated by the production at hgih energies.
 %If the reheating temperature is high enough, axinos are in thermal equilibrium and the axino yield is $\left(\frac{\nta}s\right)_{th}$. 
 %For a generic temperature, the actual yield is the lowest of the two possibilities.

In our cosmological consideration, axinos should decay before BBN, as the baryon asymmetry is generated by their decay.  
We denote by $T_D$ the axino decay temperature, defined by $\Gamma_{\tilde a}= H(T_D)$:
\beq \label{axinoTD}
T_D \simeq { 28 }\, {\rm MeV} \ |\l_{323}|\left( \frac{m_{\tilde a}}{{\rm TeV}}\right)^{1/2}
\left(\frac{ m_{\tilde a}}{m_{\tilde t_R}} \right)^2 \left(\frac{10^{12}\,{\rm GeV}}{v_{PQ}}\right).
\eeq 
The condition that the axino decays before BBN  corresponds to  $T_D\gtrsim 10 \mev$. In the absence of large axino-squark hierarchy,\footnote{
We will see below that a large splitting would give too small of a baryon asymmetry anyway.
} it gives an upper bound on $v_{PQ}$ of about $10^{12}\gev$.
Because $T_D\lesssim \mta$, axinos are non-relativistic at decay times, and  as such 
%before axinos decay they become non-relativistic when the temperature  drops below $\mta$ . 
 they will eventually dominate the energy density of the universe, unless they decay beforehand. The temperature at which the axino energy density equals the radiation energy density is
\beq
T_{eq}=\frac43 \mta\left(\frac\nta s\right)_{TP} \,.
\eeq
For $T_D>T_{eq}$, axinos decay before dominating the energy density; for $T_D<T_{eq}$, they decay after, injecting a non-negligible amount of high-energy decay products in the thermal plasma. This has the effect of  increasing the entropy, and the axino yield at decay is
\beq\label{naoversTd}
\frud{\nta}s _{T_D}=\min\left[\frac34\frac{ T_{eq}}{\mta},\frac34  \frac{ T_{D}}{\mta}\right].
\eeq
One can check that, for $T_D>T_{eq}$, the axino yield is given by Eq. \eqref{axinoyield}, while for lower decay temperatures it is given by $(3/4)(T_{D}/\mta)$.

In the axino decay, a difference in the decay rate to quarks vs. antiquarks is needed to generate a baryon asymmetry: the parameter 
$\epsilon=(\Gamma(\tilde a \to qqq)-\Gamma(\tilde a \to \bar q\bar q\bar q))/(\Gamma(\tilde a \to qqq)+\Gamma(\tilde a \to \bar q\bar q\bar q))$ gives the net asymmetry per axino decay, and the net baryon yield is
\beq
Y_{\Delta B}\equiv \frac{n_B}s=\epsilon \frud{\nta}s _{T_D}.
\eeq

On a separate note, the saxion is also produced from the thermal plasma in a similar amount as the axino. Its decay rate is much larger than that of the axino, because  the saxion can decay through the $R$-parity conserving interactions. The saxion always decays to axions with 
$\Gamma_{s\to aa} = m_s^3/(64\pi v_{PQ}^2) $, where $m_s$ is the saxion mass. 
If %the following process is
kinematically allowed, the dominant decay mode of the saxion would be $s\to hh$ with  
\beq
\Gamma_{s\to hh} &= \frac{\mu^4 }{4\pi v_{PQ}^2 m_s}\left(1- \frac{4m_h^2}{m_s^2} \right)^{1/2}
\simeq\  10^6 \left(\frac{\mu^2}{m_{\tilde a} m_s}\right)\left(\frac{m_{\tilde t_R}}{m_{\tilde a}}\right)^4
\left(\frac{\mu}{1\tev}\right)^2 \Gamma_{\tilde a}.
\eeq
Thus saxions decay much earlier than the axinos.
Furthermore, because their decays do not produce any baryon asymmetry, the role of the saxion  really is negligible.

\subsection{Axino baryogenesis}\label{subsec:Axino_baryogenesis}
As discussed in section \ref{sec:NWtheorem}, no contribution to $\epsilon$ comes at one-loop. The interference between the tree-level decay and the two-loop decay (involving the $\Delta B=2$ interactions in Fig. \ref{fig:axino_tree}(c,d)) gives a non-zero asymmetry,
\beq\label{epsasymmetry}
{\epsilon }  & {=  \epsilon_A + \epsilon_{\tilde g} +\epsilon_{\tilde B}} ,\\
{ \epsilon_A}&=\frac{|\l_{323}|^4}{{32}\pi^3}
\frac{m_t^2\mta^2}{m_{\tilde t_R}^2 m_{\tilde b_R}^2 m_{\tilde s_R}^2}{\rm Im}[\mta A''^*_{323}] 
%\left[1+\frac12\frac{\mta X_t}{m_{\tilde t_L}^2} +\frac{1}{16}\left(\frac{\mta X_t}{m_{\tilde t_L}^2}\right)^2\right], 
\frac{1 + 
\frac{1}{2}\frac{ \mta X_t}{m_{\tilde t_L}^2} 
+ \frac{1}{16}\left(\frac{\mta X_t}{m_{\tilde t_L}^2}\right)^2 }{1+\frud{X_tm_t}{m_{\tilde t_L}^2}^2}\,,
\\
{\epsilon_{\tilde g}}& { = \frac{\kappa_{\tilde g}|\l_{323}|^2g_3^2 }{32\pi^3}
\frac{m_t^2 m_{\tilde a}^2}{m_{\tilde g}^2 m_{\tilde t_R}^4}{\rm Im}[\mta m_{\tilde g}^*]
%\left[1 + 
%\frac{1}{2}\frac{ \mta X_t}{m_{\tilde t_L}^2} 
%+ \frac{1}{16}\left(\frac{\mta X_t}{m_{\tilde t_L}^2}\right)^2 \right]\,,
\frac{1 + 
\frac{1}{2}\frac{ \mta X_t}{m_{\tilde t_L}^2} 
+ \frac{1}{16}\left(\frac{\mta X_t}{m_{\tilde t_L}^2}\right)^2 }{1+\frud{X_tm_t}{m_{\tilde t_L}^2}^2}\,,
}\\
{\epsilon_{\tilde B}}& { = \frac{\kappa_{\tilde B}|\l_{323}|^2g'^2 }{32\pi^3}
\frac{m_t^2 m_{\tilde a}^2}{m_{\tilde B}^2 m_{\tilde t_R}^4}{\rm Im}[\mta m_{\tilde B}^*]
%\left[1 +  \frac{1}{2}\frac{ \mta X_t}{m_{\tilde t_L}^2} + \frac{1}{16}\left(\frac{\mta X_t}{m_{\tilde t_L}^2}\right)^2 \right]\,,
\frac{1 + 
\frac{1}{2}\frac{ \mta X_t}{m_{\tilde t_L}^2} 
+ \frac{1}{16}\left(\frac{\mta X_t}{m_{\tilde t_L}^2}\right)^2 }{1+\frud{X_tm_t}{m_{\tilde t_L}^2}^2}\,,
}
\eeq
{ 
where $\kappa_{\tilde g},\, \kappa_{\tilde B}$ are squark mass-dependent dimensionless 
parameters defined in Eq.~(\ref{kappa_gB}).
In the limit of universal squark masses, 
we get }
\beq
&{\kappa_{\tilde g}= 2,\quad \kappa_{\tilde B}= \frac{14}{9}.} 
\eeq 
Here, all mass squared terms represent real values. 
{  We can check the bino contribution would be same order of that of gluino's if the GUT relation 
($m_{\tilde B}:\, m_{\tilde g} \simeq  (3/5) g'^2:\, g_3^2$) is satisfied. In the following discussion 
we consider the case with $m_{\tilde g} \sim m_{\tilde B}$ such that $\epsilon_{\tilde g} \gg \epsilon_{\tilde B}$.}
{ As for the decay rate $\Gamma_{\tilde a}$,} this is computed in the limit of massless final states and heavy intermediate squarks, and the exact expression is discussed in the Appendix.
In the following we will denote  by $\Phi_A$ the relative phase between the axino mass and the RPV $A$-term $A''_{323}$, $\Phi_A\equiv {\rm Im}[\mta A''^*_{323}] /|\mta A''_{323}|$, 
and  by $\mtq$ the average squark mass scale,  $\mtq \equiv (m_{\tilde t_R}^2 m_{\tilde b_R}^2 m_{\tilde s_R}^2)^{1/6}$. { Similarly  
$\Phi_{\tilde \lambda} \equiv {\rm Im}[\mta m^*_{\tilde \lambda}]/|\mta m_{\tilde \lambda}| $, and 
for the contribution of $\epsilon_{\tilde g}$,  universal squark masses are taken
so that  $\mtq= m_{\tilde t_R} = m_{\tilde b_R} = m_{\tilde s_R}$.}
%{ Either $\epsilon_A$ and $\epsilon_{\tilde g}$ can be enough to reproduce the baryon asymmetry if the CP phases of each term are not small.}  
As an example, we give three benchmark points that reproduce the correct baryon asymmetry: taking $\mta=500\gev, \ A''_{323}\simeq X_t\simeq \mtq$ and {a common CP phase} ${ \Phi_A=\Phi_{\tilde g}=\Phi_{\tilde B}\equiv\Phi}$, the other parameters are:
\beq
{
{\rm BP1}:}%BP 2 in Mathematica
\qquad&\ \mtq=900\gev,\ \mtg={1.5}\tev,\ |\l_{323}|=1,\ \Phi=1;&\epsilon=3.4\times 10^{-6};
\nn\\
{ %BP 1 in Mathematica
{\rm BP2}:}\qquad&\ \mtq=1\tev,\ \mtg=2\tev,\ |\l_{323}|=0.5,\ \Phi={ 0.2}\,;&\epsilon=5.9\times 10^{-8};
\label{BP}
\\
{%BP 1.5 in Mathematica
{\rm BP3}:}\qquad&\ \mtq=2\tev,\ \mtg=1\tev,\ |\l_{323}|=1,\ \Phi={ 1}\,;&\epsilon=1.5\times 10^{-7}.
\nn
\eeq  
{ At the benchmark point BP1, the asymmetry receives roughly equal contributions from the diagrams with the $A$-terms and the gluino-mediated processes, while at BP2 and  BP3 the gluino contribution is the dominant source. This is easily explained by the large power of $\l_{323}=0.5$ for $\epsilon_A$ at BP2 and by the small gluino mass boosting $\epsilon_{\tilde g}$ at BP3.
 For each benchmark point, we obtain the observed baryon asymmetry $Y_{\Delta B}$ by choosing appropriate values for $T_R$ and $v_{PQ}$% determining $\frac \nta s$
. As these parameters also determine the dark matter abundance, we} 
 will discuss the benchmark points in relation to the nature of dark matter in the next section. 
But first, let us discuss our result, Eq. \eqref{epsasymmetry} in more details: 
\begin{itemize}

\item A large $\l_{323}$ is strongly preferred, { especially for $\epsilon_A$.} In particular, the lower limit on $\l_{323}$ is $0.03$ for $\epsilon_{\tilde g}$ to generate the asymmetry (and $0.1$ for $\epsilon_A$).
 An upper limit of $\l_{323}=1.07$ was found from the condition that perturbativity is valid up to the GUT scale % How large can it be? 
in  \cite{Allanach:1999mh}, where the RG running of the \rpving couplings was considered. %, and an absolute upper limit of $\l_{323}=1.07$ was set from the condition that perturbativity is valid up to the GUT scale. 
Although from the point of view of a low-energy effective theory this is not a problem, and one can just expect that new degrees of freedom appear around the Landau pole, we will assume $\l_{\rm max}=1$ in the rest of this paper.

\item The baryon  asymmetry is proportional to, roughly, $(\mta/\mtq)^3(m_t/\mtq)^2$. 
It is suppressed for a large hierarchy between the squark mass and either the axino mass or the weak scale. Even for ${\l_{323}=1,\,\mta\simeq\mtq \simeq m_{\tilde g}}$, there is a suppression by $m_t^2/\mtq^2$, which points to an upper limit on the squark scale.
{ In particular, in this limit we can write:
\beq\label{YYobs}
\frac{Y_{\Delta B}}{(Y_{\Delta B})}_{obs}\simeq\frud{\nta/s}{10^{-3}}_{T_D}10^4\mathcal{N}\frac{m_t^2}{\mtq^2}\frac{\mta^3}{\mtq^3} \cdot c
\eeq
where $c$ is an $\mathcal O(1)$ number determined by $\frac{A''_{323}}{\mtq},\kappa_{\tilde g}, \frac{X_t}{\mtq}$ and $\mathcal N$ takes into account the ratio between the numerical and the analytical analysis (as shown in the Appendix): for $\mta\sim\mtq\sim\mtg\sim1\tev$, we find $\mathcal N\sim 6$, while for $\mta\sim\mtq\sim\mtg\sim100\tev$ the value of $\mathcal N$ increases to about $60$ (this can be expected as in this limit the top mass is massless and the only corrections are due to the internal propagators, which increase the results).
In any case, we find 
\beq
 \mtq\lesssim 10^2 \sqrt{ \mathcal{N}c} m_t \simeq 130 \tev
 \eeq}
We find an absolute upper bound on the squark masses at {130\tev}\ for large soft terms $A''_{323}\simeq X_t\simeq 3\mtq$; larger values for $A$-terms are potentially dangerous in that they can generate color-breaking vacua \cite{Camargo-Molina:2013sta,Blinov:2013fta}.

\item Additionally, the baryon asymmetry is proportional to the relative phases between the axino mass $\mta$ and the soft SUSY breaking parameters $A''_{323}$ {and  $m_{\tilde g}$}.  
There is no direct constraint on the CP phase of $A''_{323}$ but an indirect constraint
is provided by the null results in the measurement of the neutron Electric Dipole Moment (EDM). 
CP phases for the MSSM A-terms $A_{U,D}$ (in particular, the phase $\phi_{A_Q\tilde g}\equiv {\rm Im}[\mtg A_Q^*]/\mtg^2$, where $m_{\tilde g}$ is the gluino mass and $Q=U,D$) contribute to the neutron EDM  \cite{Polchinski:1983zd},
\beq
|d_n|\simeq 2.5\times 10^{-25}e\text{ cm} \frud{\phi_{A_Q\tilde g}}{1/4}\frd{\tev}{\mtq}^2
\eeq
while experimentally the upper limit is $|d_n|<2.9\times 10^{-26} e\text{ cm} $  \cite{Baker:2006ts}. This implies either a small phase $\phi_{A_Q\tilde g}$ or superpartners in the multi-TeV range.
In our model, we can have a large baryon asymmetry and a small contribution to the neutron EDM in two ways. First, unlike the models of  \cite{Cline:1990bw} in which gluino decays contribute to the baryon asymmetry {with the same CP phase as the neutron EDM}, even with a common CP phase for all the $A$-terms  the baryon asymmetry and the neutron EDM are proportional to different phases. Thus ${\rm Arg}[\mta A''^*_{323}]={\rm Arg}[\mta A_Q^*]$ could be maximal, while Arg$[\mtg A_{Q}^*]$ could be small. 
{ In this case, ${\rm Arg}[\mta\mtg^*]$ would also be maximal, and the baryon asymmetry receives contribution from both the $A$-term and the gluino.
}
%{This possibility is also good to enhance the asymmetry by making both $\epsilon_A$ and $\epsilon_{\tilde g}$ non-negligible.} 
Second, 
the phase of $A_Q$ and $A''_{323}$ might be independent at the messenger scale so that $A''_{323}$ has a large phase while the MSSM $A$-terms could have small ones. The RG running can generates a non-zero (but small) $A_Q$ phase at low energies, that does not contribute too much to the neutron EDM.

Summarizing, the contributions to the neutron EDM depends on the SUSY breaking sector and on the phases generated at that scale. The CP phase needed for baryogenesis is not the same as the one contributing to neutron EDM.
\end{itemize}

To conclude this section, we have a mechanism for generating the right baryon asymmetry that points to large \rpv, not too large squark masses, and can be safe from the null experimental results for neutron EDM. Large \rpv\ is not a problem if it is confined in interactions involving heavy quarks, otherwise there are many potentially large baryon-number-violating  contributions to low-energy flavor physics (see \cite{Barbier:2004ez} for a review). Even if the only non-zero coupling is $\l_{323}$, couplings involving light quarks are generated at 1-loop level \cite{Dimopoulos:1987rk}: for $\l_{323}\simeq1$,  we find $\l_{112}\simeq10^{-8}, \l_{223}\simeq 10^{-5}$, which are too small to significantly contribute to $K\bar K$ mixing or $n$-$\bar n$ oscillation. We can also revisit the assumption of single-coupling dominance in the decay of the axino and see if the presence of other couplings is consistent with flavor physics. An important bound for the case with a non-negligible $\l_{313}$ coupling comes from contributions to $\Delta m_K$  \cite{Slavich:2000xm},
\beq
|\l_{323}\l^*_{313}|<3\times 10^{-2}\fru{m_{\tilde t}}{1\tev}^2
\eeq
which for $\l_{323}\simeq1$ and TeV-scale squarks, implies $|\l_{313}|\lesssim3\times 10^{-2}$. Then, the single-coupling dominance assumption was justified and  the $\l_{313}$-mediated contribution to the axino decay is negligible.

\section{Dark Matter}\label{sec:dm}
We now turn our attention to the presence of dark matter. We first recall the dark matter density from the Planck satellite's CMB measurements (combined with WMAP9 polarization maps)  \cite{Ade:2013zuv},
% combination of the Planck temperature power spectrum with a WMAP polarization low- multipole likelihood
\beq\label{DMdensity}
\Omega_{DM}h^2 = 0.1199\pm0.0027\,.
\eeq

Without \rpa, no supersymmetric particle is stable and indeed the axino, which can be a viable dark matter candidate in \rpa-conserving models  \cite{Covi:1999ty,Covi:2001nw}, decays and generates the baryon asymmetry. There are two natural candidates for dark matter that are already in  the model: axions and gravitinos. 
Coherent oscillation of the axions can give rise to cold dark matter if the PQ symmetry breaking scale is properly taken. 
If the gravitino is the LSP, its lifetime can be long enough that it constitutes the dark matter at present times. The abundance of the gravitino can be sizable by taking proper values of its mass and the reheating temperature $T_R$.  

\subsection{Heavy gravitino scenario}
When the gravitino is heavy enough to decay through the $R$-parity conserving interactions, 
the only possible candidate for dark matter is the axion.
Axion cold dark matter is generated 
when the axion starts to oscillate coherently at the QCD phase transition.
Its abundance is given as \cite{Beringer:1900zz} 
\beq\label{ADM}
\Omega_{a}h^2 = \frac{1}{\Delta_a}k_a \theta_a^2 \frud{v_{PQ}}{10^{12}\gev}^{7/6},
\eeq
where $k_a$ is a numerical factor of ${\cal O}(1)$, $\theta_a$ is the axion misalignment angle, and 
$\Delta_a$ is the possible dilution factor from entropy release when axinos decay after the axion coherent oscillation has started. In viable parameter regions, 
we find that $\Delta_a$ is just ${\cal O}(1)$.
The initial angle $\theta_a$ is not averaged out because we assume the PQ symmetry 
is broken from the inflation epoch.  
There is no dark matter contribution from the axionic string decays for the same reason. 
With the natural value of the angle $\theta_a^2= \langle \theta_a^2\rangle \sim 3$, $v_{PQ} \sim 10^{11}\, {\rm GeV}$  explains the present density of dark matter. From the dark matter constraint, a larger value of  $v_{PQ}$ is allowed if we take  a small value of $\theta_a$. However, $v_{PQ}$ cannot be too large, otherwise axinos  will decay after the BBN era.  %Such a high PQ scale determines weak interactions between the axino and the squarks, so that the axino production at reheating is rather suppressed (for example, the axinos are never in thermal equilibrium). As a result, it is harder to produce enough baryon asymmetry. The upper bound on the squarks mass decreases to $\mtq\lesssim4\tev$, and both large $\Phi$ and large mixing $X_t$ are preferred.
{
Using 
Eqs. (\ref{axinoTD}) and (\ref{epsasymmetry}), $v_{PQ}$ can be represented as}
\beq
{ v_{PQ} =10^{11}\gev |\lambda_{323}''|(c \Phi)^{1/2}
\left(\frac{\mta}{0.5\tev}\right)^{3/2}
\left(\frac{0.8\times 10^{-10}}{Y_{\Delta B}}\right)^{1/2}
\left(\frac{1\tev}{M_{\rm SUSY}}\right)^{5/2} 
\left(\frac{T_R}{10^7\gev}\right)^{1/2} ,}
\eeq
{ for $T_D > T_{eq}$. Otherwise, for $T_D < T_{eq}$,}
\bea
{v_{PQ} =  10^{12}\gev\left(\frac{|\lambda''_{323}|}{(c\Phi)^2}\right)^{1/5}
\left(\frac{m_{\tilde a}}{0.5\tev}\right)^{17/10}
\left(\frac{Y_{\Delta B}}{0.8\times 10^{-10}}\right)^{2/5} 
\left(\frac{10\mev}{T_D}\right)^{7/5} ,}
\eea
{ where $c$ is ${\cal O}(1)$ coefficient. 
In these expressions, we set all SUSY breaking parameters as a common scale, $M_{\rm SUSY}$, 
for simplicity.} 
Thus, we get $v_{PQ}\lesssim 10^{12}\gev$ for reasonable parameter values. 
The allowed range is rather small,
\beq
10^{11}\gev \lesssim v_{PQ} \lesssim 10^{12}\gev.
\eeq

In order to produce sizable baryon asymmetry, the reheating temperature should be high enough, but it is notable that $T_R$ need not be as large as $v_{PQ}$. This is consistent with the assumption that the PQ symmetry is not restored in the reheating epoch.  As an example, the observed dark matter abundance and baryon asymmetry are generated  for $v_{PQ}= 10^{11}\gev, \, T_R=1.5\times 10^7\gev$ at the benchmark point BP1, where $\mta=500\gev,\, \mtq={900\gev},\mtg=1.5\tev,\,A''_{323}\simeq X_t = \mtq,\, \l_{323}=1,\, \Phi={1}$. %\comment{check after dilution taken into account}
Because the baryon asymmetry is inversely proportional to the squark mass $\mtq$ and the PQ scale $v_{PQ}$ {(through the axino abundance $\nta/s$)}, with such a high value of $v_{PQ}$ we {can repeat the argument leading to Eq. \eqref{YYobs} and find an absolute upper bound of {$\mtq\lesssim 10 \sqrt {Nc} \simeq 15 \tev$}.} Note that this upper limit is found taking $\mta\simeq\mtq$ and large $A$-terms, $A''_{323}\simeq X_t\simeq 3\mtq$, so that it corresponds to a rather compressed region of the parameter space. For a more natural choice of parameters, the squark mass has to be below {8 TeV}.

On the other hand, although the gravitino is not a present dark matter candidate, its lifetime can be long enough to cause problems. 
%If the gravitino is not the true LSP (making the axino an NLSP), the axion can be dark matter; in this case, the usual bounds \cite{Moroi:1993mb} in the $\mtr-T_R$ plane can be applied, as the gravitino will decay through \rpcing interactions and might interfere with BBN or dilute the baryon asymmetry generated by axino decays. If the gravitino is effectively produced at reheating, we  will require $\mtr\gtrsim O(100)\tev$, corresponding to an anomaly-mediated type of SUSY spectrum. Otherwise, if the reheating temperature is low enough, the gravitino decay products are negligible.
The decay rate of the gravitino is  %\comment{reference?}
\bea
\Gamma_{3/2} =  \frac{1}{32\pi}\left(n_V +\frac{n_C}{12}\right)\frac{m_{3/2}^3}{M_P^2}
= \left(4\times 10^5\sec\right)^{-1}\left(n_V +\frac{n_C}{12}\right) \left(\frac{m_{3/2}}{{\rm TeV}}\right)^3,
\eea
where $n_V$ ($n_C$) is the number of vector (chiral) supermultiplets whose masses are smaller than the gravitino mass.   
When the gravitino is heavier than the MSSM sparticles, its decay products and their amounts are strongly constrained by successful prediction of the standard Big Bang nucleosynthesis \cite{Kawasaki:2004qu}. 
For reheating temperatures around $10^7\gev$ (needed to generate enough axinos),  the gravitinos have to decay before the BBN era, i.e. $\tau\th < {\cal O}(0.1)\sec$. 
This requires $m_{3/2} \gtrsim 50\,{\rm TeV}$, corresponding to a spectrum typical of anomaly-mediation of SUSY breaking.
%\cite{Kawasaki:2004qu}
%\bibitem{Kawasaki:2004qu} 
%  M.~Kawasaki, K.~Kohri and T.~Moroi,
  %``Big-Bang nucleosynthesis and hadronic decay of long-lived massive particles,''
% Phys.\ Rev.\ D {\bf 71}, 083502 (2005)
%  [astro-ph/0408426].
  %%CITATION = ASTRO-PH/0408426;%%
  %545 citations counted in INSPIRE as of 08 Dec 2014

The late time decay of heavy gravitinos could also contribute to the baryon asymmetry, as in \cite{Cline:1990bw}. However, in \cite{Cline:1990bw} the gravitinos were dominating the energy density of the universe at decay time, implying $T_R\sim10^{15}\gev$, while in our case the yield of the gravitino is too small to contribute to a baryon asymmetry of $n_B/s  \sim 10^{-10}$.

\subsection{Light gravitino scenario}
When the gravitino is the true LSP, the axino is the NLSP and can decay to the gravitino and the axion with a  decay rate \cite{Chun:1993vz}
\bea
\Gamma_{\tilde a\to a \psi_{3/2}}=\frac1{96\pi}\frac{\mta^5}{M_P^2\mtr^2}\left(1-\frac{\mtr^2}{\mta^2}\right)^{1/2}.
\eea
Because the branching ratio ${\rm Br}(\tilde a\to a\psi_{3/2})$ is quite small, the baryogenesis mechanism is effectively the same as for an axino LSP, and our previous discussion holds.
However the non-thermal production of gravitinos by axino decays can provide a sizable abundance of dark matter as
\beq
\Omega^{NTP}\th h^2 &=0.274 \times  10^9\,
{\rm Br}(\tilde a\to a\psi_{3/2})\left(\frac{m_{3/2}}{{\rm GeV}}\right)
\left(\frac{\nta}{s}\right)_{T_D}\nonumber\\
&= {\left(\frac{1}{\lambda''_{323}}\right)^2}
\left(\frac{\mtq}{1\tev}\right)^4\left(\frac{1\gev}{m_{3/2}}\right)
{\rm min}
\left[{0.048}\left(\frac{T_R}{10^7\,{\rm GeV}}\right)
,\,  { 0.015}\left(\frac{v_{PQ}}{10^{10}\gev}\right)^2\right].
\eeq
The second line is evaluated for $T_D > T_{eq}$. For $T_D < T_{eq}$, there is a further dilution by the factor $T_D/T_{eq}$. 
On the other hand, 
the thermal production at reheating reads \cite{Bolz:2000fu,Pradler:2006qh,Rychkov:2007uq}
\beq\label{GDM}
\Omega\th^{TP} h^2\simeq {0.07}\frud{\mtg}{1\tev}^2\frud{1\gev}{\mtr}\frud{T_R}{10^7\gev}\,.
\eeq
{ We note that for given $m_{3/2}$ and $\mtg\gtrsim \mtq$, 
the thermal production is always the dominant contribution.}
 For relatively low $T_R$, 
the non-thermal production also can be important {when the gluino is lighter than squarks, 
also for small $\lambda''_{323}$. }

If light gravitinos are produced in the right amount, they can give the correct relic density, provided that their lifetime is longer than the age of the universe (they decay via RPV interactions, $\psi\th\to qqq$). As a matter of fact, the condition on the gravitino lifetime is stronger, as the hadronic decay pro¨ducts would contribute to the cosmic ray antiparticle population, which is looked at in experiments such as PAMELA or AMS-02  \cite{Bomark:2009zm,Dal:2014nda,Monteux:2014tia}. For example, in  \cite{Ibarra:2012cc} it was shown that the lifetime of a vanilla DM candidate decaying to $b\bar b$ is constrained to be bigger than about $5\times 10^{27} \sec$ from the non-observation by PAMELA of an excess in the $\bar p/p$ fraction, for $80\gev \lesssim m_{DM}\lesssim 500\gev$ (future antideuterons experiments will do better in the lower mass range). To translate these results to the case of a gravitino decaying to three quarks, it is necessary to find how many antiprotons are generated and compare it to the case of a $b\bar b$ final state, for each value of the DM mass. This effort is being tackled by one of the authors in  \cite{CCM:inprogress}, and it is generally found that the number of antiprotons in the experimental energy range produced in the $\psi_{3/2} \to qqq$ case is approximately the same as in the $\chi\to b\bar b$ case, with variations of around $\pm 30\%$, depending on the particle mass and the specific flavor structure of the final three-quarks state. It is then reasonable to take the lower bound $\tau_{\psi_{3/2}}^{\rm exp}\gtrsim10^{27} \sec$ on the gravitino lifetime, when it makes up all of the dark matter. This bound is conservative enough to not be sensitive to the uncertainties in the precise number of antiparticles arising from the gravitino decay.

The gravitino lifetime can be computed as \cite{Monteux:2014tia}
\beq
\label{32lifetime}
\tau_{\psi_{3/2}\to u_id_jd_k}=1.28\times 10^{26} \text{sec}\left(\frac1{\lambda''_{ijk}}\right)^2 \left(\frac{\text{3 GeV}}{\mtr}\right)^7\fru{\mtq}{1\tev}^4.
\eeq
In our model the biggest coupling is $\l_{323}$, allowing the decay channel $\psi_{3/2}\to tbs$ for $\mtr\gtrsim m_t+m_b+m_s$. For a gravitino lighter than the top quark, decays would be mediated by the biggest $\l_{ijk}$ coupling with $i\neq3$. For example, if the next non-negligible coupling is $\l_{223}$ the decay would go through the $cbs$ channel down to the bottom quark mass. Even with $\l_{223}\simeq1$, this coupling would not contribute to  baryogenesis as the axino partial decay rate would be proportional to the charm quark mass (instead of the top quark mass). Remembering that $\l_{223}\simeq10^{-5}\l_{323}$ is generated at 1-loop anyway, we can consider the range $10^{-5}\lesssim \l_{223} \lesssim 1$. The lower bound on the DM lifetime $\tau\th\gtrsim 10^{27}$ seconds implies an upper bound on the gravitino mass, $60 \gev \gtrsim  \mtr^{\rm max}\gtrsim 4\gev $ (for $\mtq=1\tev$; these bounds scale as $\mtq^{4/7}$).

Finally, there is also a lower bound on the gravitino mass, coming from the one-loop proton decay channel $p\to K^+\psi_{3/2}$ setting a limit on $\l_{323}$  \cite{Choi:1998ak}: 
\beq
\l_{323}\leq5\times 10^{-8}\fru{\mtq}{300\gev}^2\fru{\mtr}{1\text{ eV}}
\eeq
For $\l_{323}=1,\ \mtq=1\tev$, the corresponding lower limit on the gravitino mass is $\mtr\gtrsim 2 \mev$.

\begin{figure}[t]
\begin{center}
\subfigure[]{\includegraphics[width=0.465\textwidth]{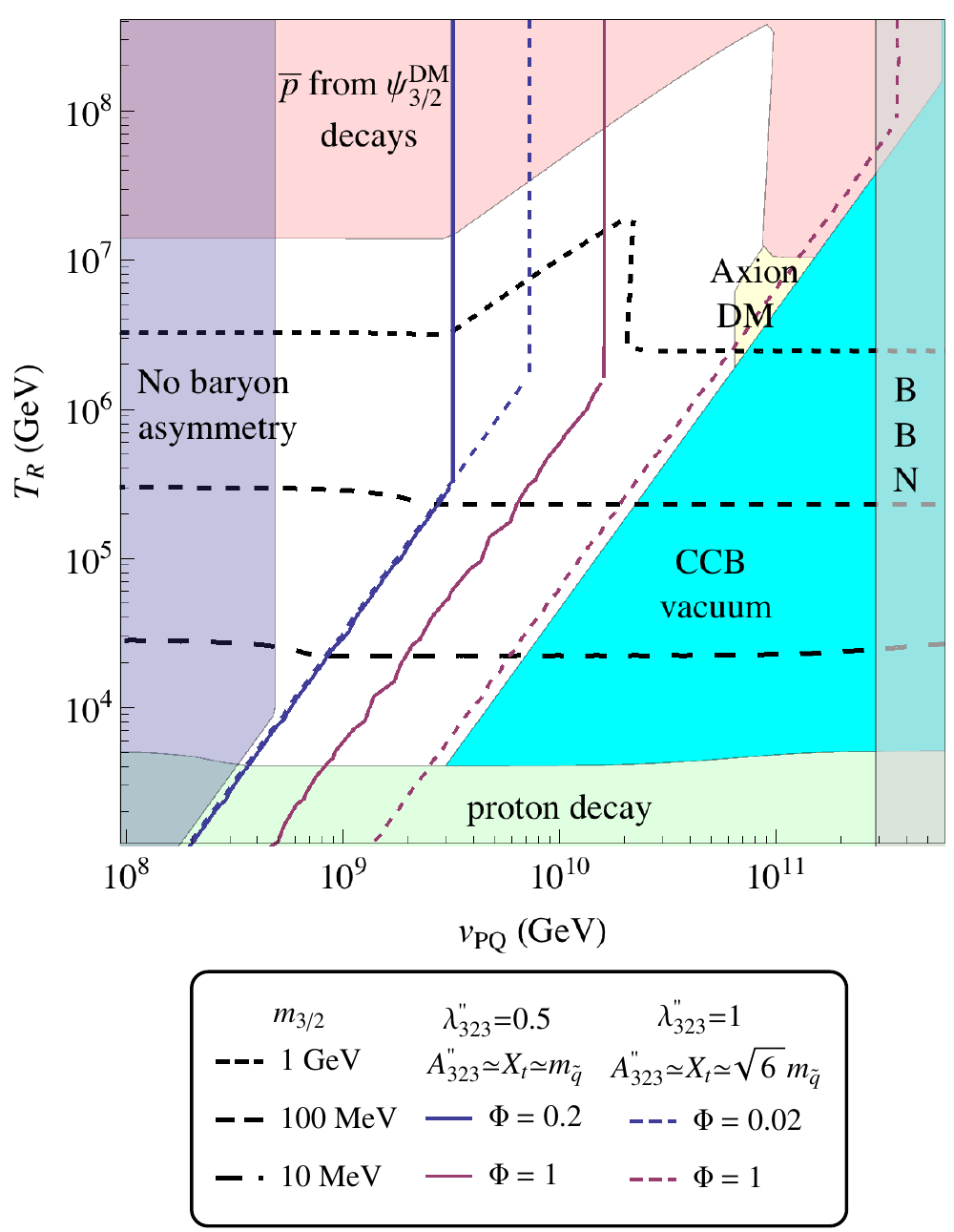}}\quad
\subfigure[]{\includegraphics[width=0.46\textwidth]{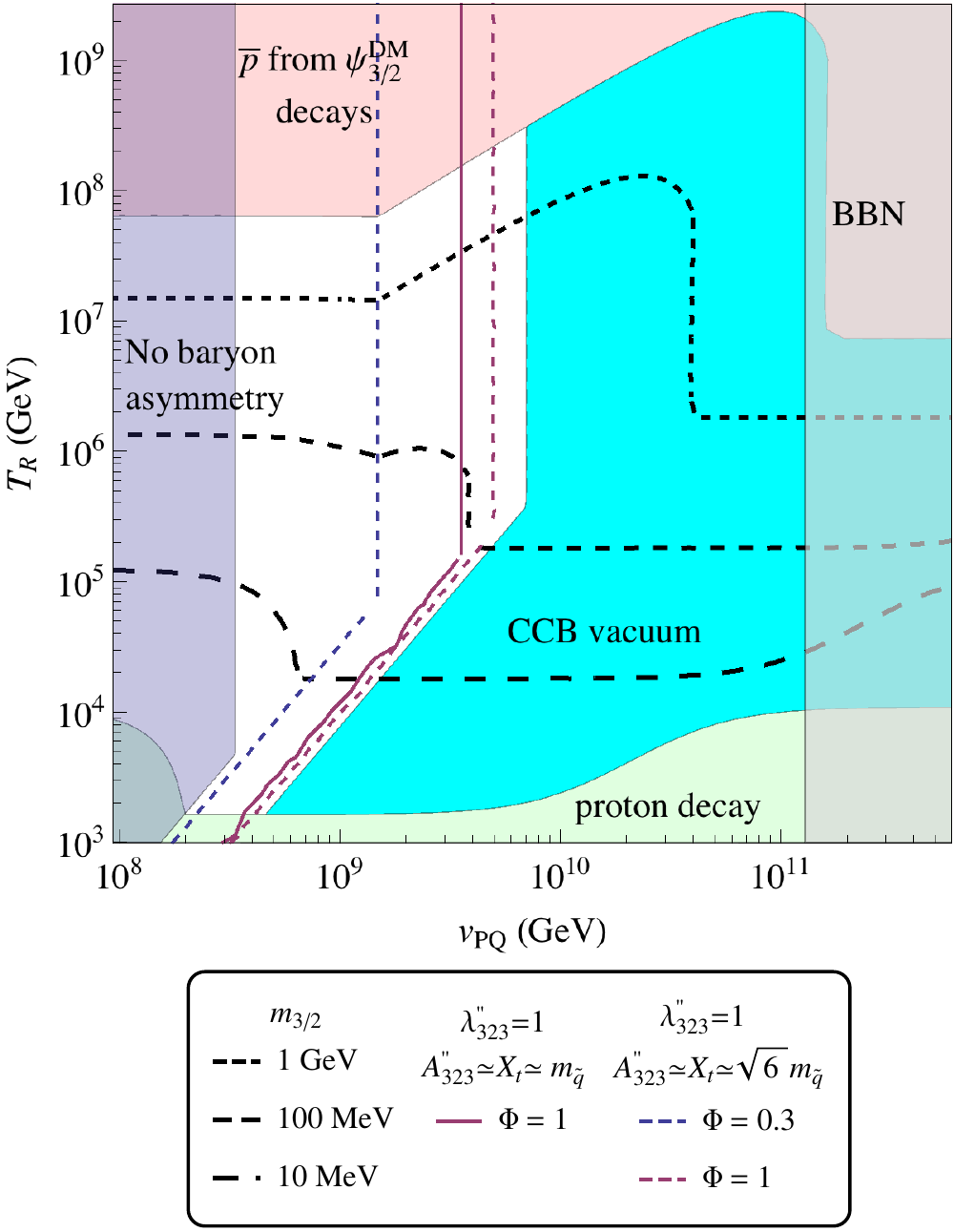}}
\caption{
{Constraints in the $T_R-v_{PQ}$ plane, keeping $\mta=500\gev$ fixed. The black dashed lines are contours for different gravitino masses that give the observed dark matter abundance, while the diagonal/vertical continuous and dashed lines correspond to different values of $\Phi_g=\Phi_A=\Phi, A_{323}'', X_t$ reproducing the correct baryon asymmetry. The shaded regions are excluded by the corresponding labelled constraints, which are further explained in the text. { (a)} Masses are fixed at $\mtq=1\tev,\,\mtg=2\tev$, corresponding to the benchmark point BP2.  { (b)} for a light gluino, the baryon asymmetry is dominated by the gluino-mediated diagram; the fixed parameters are $\mtq=2\tev,\, \mtg=1\tev$, corresponding to BP3.
}
%Parameter space constraints in the $T_R-v_{PQ}$ plane, with fixed $\mta=500\gev,\mtq=1\tev,\l_{323}=1$. The black dashed lines are contours for different gravitino masses that give the observed dark matter abundance, while the diagonal/vertical continuous and dashed lines correspond to different values of $\Phi, A_{323}'', X_t$ reproducing the correct baryon asymmetry. The shaded regions are excluded by the corresponding labelled constraints, which are further explained in the text.
}
\label{fig:gravDM}
\end{center}
\end{figure}

In Fig. \ref{fig:gravDM}, we fix %the RPV coupling to  $\l_{323}=1$, 
the axino %and squark masses
mass
to $\mta=0.5\tev$ % and $\mtq=1\tev$, 
and vary the remaining parameters ($\l_{323},A''_{323},X_t, \mtr$) in the $v_{PQ}-T_R$ plane. The black horizontal dashed lines are contours of different values of $\mtr$ that give the correct dark matter relic abundance in the range $2\mev \lesssim\mtr \lesssim 4\gev$. The ranges excluded by proton decay and cosmic ray observations are respectively shown  at the bottom in green and at the top in red (taking $\l_{223}\simeq1$). The non-thermal production from axino decays contributes to a dip in the lines, more easily seen in the figure on the right. For %higher gravitino masses (and
 higher reheating temperatures, %)
  a spike can be seen when axinos decay  as they dominate the energy density of the universe ($T_D<T_{eq}$: at higher reheating temperatures axinos are produced more efficiently); this dilutes the gravitinos produced at reheating, allowing a higher reheating temperature than naively thought.%, resulting in a higher number of produced gravitinos.
The diagonal lines (becoming vertical around the center of the plot) are contours that give the correct baryon asymmetry, with different values of the soft terms and the CP phase $\Phi$ %={\rm Im}[\mta A''^*_{323}]/|\mta A''_{323}|$
. Their behavior can be understood in the following way: for low reheating temperatures, the axino yield depends on both $T_R$ and on the PQ scale $v_{PQ}$, see Eq. \eqref{axinoyield}. For high $T_R$, the yield is just given by the thermal scattering expression, independent of $v_{PQ}$. As we increase the variables $\Phi,A'',X_t$ that determine the asymmetry parameter $\epsilon$, the correct baryon asymmetry can be generated at higher values of $v_{PQ}$, that is, with weaker axino interactions. In the shaded region to the left, not enough asymmetry can be generated, because the axino yield is too small to start with and $\epsilon$ cannot be too large. In the light-blue region on the right, a correct baryon asymmetry can be generated only by taking large values of the soft $A$-terms, which is dangerous from the point of view of color-breaking vacua (the stop squarks might acquire a vev). We excluded the region with $A''_{323}\simeq X_t\gtrsim 3\mtq$.\footnote{This is a conservative estimate, as slightly lower values of $X_t$ could also generate unstable/metastable vacua. See Refs. \cite{Camargo-Molina:2013sta,Blinov:2013fta} for a more detailed discussion on the constraints on the MSSM $A$-terms. More recently, RPV $A$-terms were considered in Ref. \cite{Chamoun:2014eda}, under the assumption of CMSSM-like boundary conditions (universal $A$-terms at the GUT scale). Because the masses of $\tilde b_L$ and $\tilde s_L$  are also important for vacuum stability but do not appear in the baryon asymmetry parameter $\epsilon$ in Eq. \eqref{epsasymmetry}, we leave the general study of the effect of RPV $A$-terms for future work.
}
On the right, in the light yellow region the axion can be dark matter (depending on the precise value of the misalignment angle) and gravitinos can either have decayed already or be a sub-dominant dark matter component, while in the gray rightmost region the axino decays at $0.1-1\sec$, compromising the observed abundances for light nuclei produced during BBN.

We note that both baryogenesis and dark matter can be accounted for in most of the ``axion window'', $10^9\gev<v_{PQ}<10^{12}\gev$, for reheating temperatures as low as 10\tev\, and as high as $10^7-10^8\gev$. 
It is interesting to point out that the choice of fixed parameters in Fig. \ref{fig:gravDM} is in some way optimal: for heavier squark masses (for fixed $\mta/\mtq$) the asymmetry parameter $\epsilon$ becomes smaller. For an almost  degenerate axino LSP, $\mta\approx \mtq$, more parameter space opens up, as $\epsilon$ is bigger, and smaller $A$-terms (and phases) are allowed; in this case, the higher squark mass allowed is {35\tev.}

We finish this sub-section with a comment on the Higgs mass:  large $A$-terms are needed to achieve a 125 \gev\ Higgs boson with light stops in the MSSM, and at the same time large $A$-terms increase the asymmetry parameter $\epsilon$. A 125\gev\ Higgs with maximal mixing ($X_t\simeq \sqrt6\mtq$) allows non-maximal values for $\l_{323}$ and the CP-violating phase $\Phi$, such as $\Phi=.03$.

\section{Conclusions}\label{sec:conclusions}

We have discussed a new mechanism for baryogenesis through the \rpving decay of an axino LSP, at the two-loop level and at the second order in the \bnving couplings. A suitable dark matter density is also generated by related processes, namely  by the coherent oscillation mechanism for axions and by thermal scatterings and  the axino decays for gravitinos. The scenarios described are very predictive: for the case of axion dark matter, the allowed range for the squarks extends to about {15 TeV}; additionally, the initial axion misalignment angle is large. For the case of gravitino dark matter, the gravitino mass is between a few \mev\ and a few \gev, with proton decay and cosmic rays experiments capable of narrowing this interval; in this case the upper limit on squark masses is higher, of order {$130\tev$}. The cited limits on the squark masses correspond to tuned regions of the parameter space, where the axino mass is very close to the squarks masses; requiring that the axino and the squarks masses are different by at least 20\% brings down the upper squark mass limits to { 8 TeV and 90 TeV}.
In both cases, the axino should be close to the squark mass, up to a factor of a few, and \rpv\ should be maximal, corresponding to prompt decays of superpartners. 

At the LHC, the most important signatures of light RPV squarks are multijets, with at least two jets from each squark, and three jets from the decay of a gluino. The most relevant experimental searches are \cite{Chatrchyan:2012uxa} from CMS and \cite{ATLAS-CONF-2013-091,ATLAS:2012dp} from ATLAS. In particular, Ref. \cite{ATLAS-CONF-2013-091} studied the decay of pair-produced gluinos to six quarks, and used $b$-tagging to probe the flavor structure of the RPV couplings $\l_{ijk}$. Gluino masses below $874\gev$ are excluded for gluinos whose decay products include a top and a bottom (as it is the case for large $\l_{323}$ coupling). {This limits are close to excluding the benchmark point BP3, which had $\mtg=1\tev$.} Unfortunately the limits on the gluino masses, apart from the matter of naturalness, are of little importance for our baryogenesis model, { which can be mediated by squarks only (even for the gluino-mediated process, the dependence on $\mtg$ is weak, $\epsilon_{\tilde g}\propto1/\mtg$). }
In fact, because the cross section for pair-produced stops is smaller than for pair-produced gluinos, RPV squarks are relatively unprobed at the LHC; for example, LSP squarks are best probed at the Tevatron by the CDF experiment,  excluding squark masses up to about 100 GeV \cite{Aaltonen:2013hya}. With dedicated searches, the LHC at 14 \tev\,  has the potential to exclude RPV squarks up to about a TeV \cite{Bai:2013xla,Duggan:2013yna}. For our axion dark matter scenario, a big part of the {\it natural} region of the parameter space can be probed.

\section*{Acknowledgments}
A.M. and C.S.S. are supported in part by DOE grants doe-sc0010008, 
DOE-ARRA- SC0003883, and DOE-DE-SC0007897

\appendix
%\section{Exact results in the effective theory}
\section{${\cal L}_{|\Delta B|=2}^{\tilde \lambda}$ for non-universal right-handed squark masses}
From the Lagrangian (\ref{DeltaB2_gaugino}), we get the following 
interaction terms for the gluino exchange 
\beq
{\cal L}_{|\Delta B|=2}^{\tilde g} =
-\sum_{{\rm all\ indices}}&\frac{g_3^2 m_{\tilde g}^* 
\lambda_{\alpha\beta\gamma}\lambda_{\delta\epsilon\zeta}
\epsilon_{ijk}\epsilon_{i'j'k'}}{4|m_{\tilde g}|^2} \times \nn \\
&\left[\frac{((d^c_\beta)^j(d^c_\gamma)^k)((u^c_\alpha)^{i'}(u^c_\delta)^i 
- \frac{1}{3}(u^c_\alpha)^i(u^c_\delta)^{i'})((d^c_\epsilon)^j(d^c_\zeta)^k)
}{4 m_{\tilde u_{R\alpha}}^2 m_{\tilde u_{R\delta}}^2} \right.  \nn\\
&\ +  \frac{((u^c_\alpha)^i(d^c_\beta)^j)((d^c_\gamma)^{k'}(u^c_\delta)^k 
- \frac{1}{3}(d^c_\gamma)^k (u^c_\delta)^{k'})((d^c_\epsilon)^{i'}(d^c_\zeta)^{j'})}{
m_{\tilde d_{R\gamma}}^2  m_{\tilde u_{R\delta}}^2 } \nn\\
&\ \left.+ \frac{((u^c_\alpha)^i(d^c_\beta)^j)((d^c_\gamma)^{k'}(d^c_\zeta)^k 
- \frac{1}{3}(d^c_\gamma)^k(d^c_\zeta)^{k'})((u^c_{\delta})^{i'}(d^c_{\epsilon})^{j'})
}{ m_{\tilde d_{R\gamma}}^2 m_{\tilde d_{R\zeta}}^2} \right],
\eeq 
and for the bino exchange
\beq
{\cal L}_{|\Delta B|=2}^{\tilde B} =
-\sum_{{\rm all\ indices}}&\frac{g'^2 m_{\tilde B}^* 
\lambda_{\alpha\beta\gamma}\lambda_{\delta\epsilon\zeta}
\epsilon_{ijk}\epsilon_{i'j'k'}}{2|m_{\tilde B}|^2} \left[ 
\frac{ Y_{u^c}^2((d^c_\beta)^j(d^c_\gamma)^k)((u^c_\alpha)^i(u^c_\delta)^{i'})((d^c_\epsilon)^j(d^c_\zeta)^k)
}{4 m_{\tilde u_{R\alpha}}^2 m_{\tilde u_{R\delta}}^2} \right.  \nn\\
&\  + \frac{Y_{u^c} Y_{d^c}((u^c_\alpha)^i(d^c_\beta)^j)( (d^c_\gamma)^k (u^c_\delta)^{k'})((d^c_\epsilon)^{i'}(d^c_\zeta)^{j'})}{ m_{\tilde d_{R\gamma}}^2 m_{\tilde u_{R\delta}}^2 } \nn\\
&\ \left.+ \frac{ Y_{d^c}^2((u^c_\alpha)^i(d^c_\beta)^j)((d^c_\gamma)^k(d^c_\zeta)^{k'})((u^c_{\delta})^{i'}(d^c_{\epsilon})^{j'})
}{ m_{\tilde d_{R\gamma}}^2 m_{\tilde d_{R\zeta}}^2}   \right],
\eeq
where $ijk\, i'j'k'$ are the  indices of SU(3) anti-fundamental representation, and $Y_q$ is the $U(1)_Y$ hypercharge of $q$. 
We used 
\beq
\sum_{A} \left({[T^A]^a}_b\, (q_1)^b\right)\left({[T^A]^c}_d\, (q_2)^d\right)  &= 
\frac{1}{2} (q_1)^c (q_2)^a - \frac{1}{6} (q_1)^a (q_2)^c,\eeq
for SU(3)$_c$ gauge group. If the coupling $\lambda_{323}''= - \lambda_{332}''$ is only nonzero, those interactions become
\beq
{\cal L}_{|\Delta B| =2}^{\tilde g} 
=& -\frac{g_3^2 m_{\tilde g}^* \lambda_{323}''^2 \epsilon_{ijk} \epsilon_{i'j'k'}}{3|m_{\tilde g}|^2}
\left [\left(\frac{1}{m_{\tilde b_R}^4} +\frac{1}{m_{\tilde s_R}^4} 
+ \frac{1}{m_{\tilde s_R}^2 m_{\tilde b_R}^2}  \right)((t^c)^i (b^c)^j)((s^c)^k (b^c)^{j'})((s^c)^{k'}
(t^c)^{i'} ) \right. \nn \\
&+ \left( \frac{1}{m_{\tilde t_R}^4} + \frac{1}{m_{\tilde s_R}^4}
+ \frac{1}{m_{\tilde s_R}^2 m_{\tilde t_R}^2}\right)
((b^c)^j(t^c)^i )((s^c)^k (t^c)^{i'})((b^c)^{j'}(s^c)^{k'}) \nn\\
&\left.+ \left(\frac{1}{m_{\tilde b_R}^4} + \frac{1}{m_{\tilde t_R}^4} 
+ \frac{1}{m_{\tilde b_R}^2 m_{\tilde t_R}^2} \right) ((t^c)^i (s^c)^k)((b^c)^j (t^c)^{i'})((b^c)^{j'}(s^c)^{k'} )\right]
\eeq
and 
\beq
{\cal L}_{|\Delta B|=2}^{\tilde B} = & -\frac{g'^2 m_{\tilde B}^* \lambda_{323}''^2 \epsilon_{ijk} \epsilon_{i'j'k'}}{9|m_{\tilde B}|^2}
\left [\left(\frac{1}{m_{\tilde s_R}^4} 
+\frac{1}{m_{\tilde b_R}^4} 
- \frac{2}{m_{\tilde s_R}^2 m_{\tilde b_R}^2}  \right)((t^c)^i (b^c)^j)((s^c)^k (b^c)^{j'})((s^c)^{k'}
(t^c)^{i'} ) \right. \nn \\
&+ \left( \frac{4}{m_{\tilde t_R}^4} + \frac{1}{m_{\tilde s_R}^4} 
+ \frac{2}{m_{\tilde s_R}^2 m_{\tilde t_R}^2}\right)
((b^c)^j(t^c)^i )((s^c)^k (t^c)^{i'})((b^c)^{j'}(s^c)^{k'}) \nn\\
&\left.+ \left( \frac{4}{m_{\tilde t_R}^4} + \frac{1}{m_{\tilde s_R}^4} 
+ \frac{2}{m_{\tilde b_R}^2 m_{\tilde t_R}^2} \right) ((t^c)^i (s^c)^k)((b^c)^j (t^c)^{i'})((b^c)^{j'}(s^c)^{k'} )\right].\eeq
We used the identity 
$(\psi_x \psi_y)(\psi_z \psi)+(\psi_y\psi_z)(\psi_x \psi) + (\psi_z\psi_x)(\psi_y \psi)=0$ 
for chiral fermions $\psi_x,\, \psi_y,\, \psi_z,\,\psi$.
Now we can easily evaluate $\epsilon_{\tilde g}$ and $\epsilon_{\tilde B}$, 
and obtain the result of (\ref{epsasymmetry}) with  
\beq \label{kappa_gB}
\kappa_{\tilde g} = &\frac{1}{3} + \frac{m_{\tilde t_R}^4}{6m_{\tilde s_R}^4} + 
\frac{m_{\tilde t_R}^4}{6m_{\tilde b_R}^4}  + \frac{m_{\tilde t_R}^2}{6m_{\tilde s_R}^2} 
+ \frac{ m_{\tilde t_R}^2}{6m_{\tilde b_R}^2}\,,\nn\\
\kappa_{\tilde B} = &\frac{8}{9} + \frac{m_{\tilde t_R}^4}{9m_{\tilde s_R}^4} + 
\frac{m_{\tilde t_R}^4}{9m_{\tilde b_R}^4}  + \frac{2m_{\tilde t_R}^2}{9m_{\tilde s_R}^2} 
+ \frac{2m_{\tilde t_R}^2}{9m_{\tilde b_R}^2}\,.  
\eeq

\section{Exact expressions for decay rate and asymmetry parameter}

In section \ref{sec:baryogenesis} we presented the decay rate $\Gamma_{\tilde a}$ and the asymmetry parameters $\epsilon_A,\epsilon_{\tilde g}$ in the limit of heavy internal squarks and massless final states. 
From the Feynman diagrams in Fig.~\ref{fig:axino_tree}, 
the full expressions are of the form
\beq
\Gamma_{\tilde a\to t^\alpha b^\beta s^\gamma}=& \varepsilon^{\alpha\beta\gamma}\varepsilon_{\alpha\beta\gamma}\frac{\kappa_3^2m_t^2|\l_{323}|^2}{2\mta v_{PQ}^2 }
\int \frac{d^3 {\bf p}_t}{(2\pi)^32p_{t0}} \frac{d^3 {\bf p}_b}{(2\pi)^32p_{b0}}\frac{d^3 
{\bf p}_s}{(2\pi)^32p_{s0}} (2\pi)^4\delta^{(4)}(p_I-p_t-p_b-p_s)
\nn\\
&\times \frac{4\left((p_I\cdot p_t)(p_b\cdot p_s)+ 2m_t\mta (p_b\cdot p_s) \frac{ m_tX_t}{m_{\tilde t_L}^2}+(p_I\cdot p_t)(p_b\cdot p_s)\frac{ m_t^2X_t^2}{m_{\tilde t_L}^4}\right)}{
((p_I - p_t)^2 - m_{\tilde t_R}^2)^2},
\eeq
\beq
\epsilon_A=&\frac{-6\kappa_3^2 m_t^4|\lambda_{323}''|^6  {\rm Im}[ m_{\tilde a}A_{323}''^*]}{m_{\tilde a}v_{PQ}^2\Gamma_{\tilde a} }
\int \frac{d^3 {\bf p}_t}{(2\pi)^32p_{t0}} \frac{d^3 {\bf p}_b}{(2\pi)^32p_{b0}}\frac{d^3 {\bf p}_s}{(2\pi)^32p_{s0}}  (2\pi)^4\delta^{(4)}(p_I-p_t-p_b-p_s)
%\varepsilon^{\alpha\beta\gamma}\varepsilon_{\alpha\beta\gamma}\frac{|\l_{323}|^4m_t^2}{}
\nn\\
&\int \frac{d^3 {\bf k}_t}{(2\pi)^32k_{t0}} \frac{d^3 {\bf k}_b}{(2\pi)^32k_{b0}}\frac{d^3 {\bf k}_s}{(2\pi)^32k_{s0}} 
(2\pi)^4\delta^{(4)}(p_I-k_t-k_b-k_s) 
\times \nn\\
&\frac{4(p_b\cdot p_s)(k_b\cdot k_s) \left(1+ 2(p_I\cdot p_t) \frac{ X_t}{ m_{\tilde a}m_{\tilde t_L}^2}+(k_t\cdot p_t)\frac{ X_t^2}{m_{\tilde t_L}^4}\right)
}
{ ((k_b+k_s)^2-m_{\tilde t_R}^2)
 ((p_b+p_s)^2-m_{\tilde t_R}^2 )
  ((k_b+k_t)^2-m_{\tilde s_R}^2)
   ((p_t-k_s)^2-m_{\tilde b_R}^2 )
  ((p_b+p_s)^2-m_{\tilde t_R}^2 )
}
\eeq
where $p_I$ is the initial momentum of the axino, $p_i$'s are the momenta of the final states $t,b,s$ and $k_j$'s are the momenta of the on-shell intermediate states that generate an imaginary part for the integral.
The expression for $\epsilon_{\tilde \lambda}$ has a similar form, where the denominator is changed to include the gaugino propagator.

%If one defines coeffcieintes $z_t,z_b$, the integration region is
%\beq
%x^2< z_t <1+x^2\\
%\frac12(z_t-\sqrt{z_t^2-4x^2})<z_b<\frac12(z_t+\sqrt{z_t^2-4x^2})
%\eeq

\paragraph{Corrections for sizable top mass} 
The top quark mass is not negligible for axino masses below a TeV.
For a non-zero top mass, the available phase space is reduced, and 
we write down the full dependence on $m_t$:
\beq
\Gamma_{\rm exact}^{\rm EFT}&=\frac{\kappa_3^2 3 |\lambda_{323}|^2}{512\pi^3}\frac{m_{\tilde a}^5 m_t^2}{v_{PQ}^2 m_{\tilde t^c}^4}
 \left[f(x)+8{x g(x)}\frac{m_t X_t}{ m_{\tilde t}^2}+f(x)\left(\frac{m_t X_t}{ m_{\tilde t}^2}\right)^2
\right]\,,
\\
\epsilon_{\rm exact}^{\rm EFT}&=
\frac{|\l_{323}|^4}{32\pi^3}
\frac{m_t^2\mta^2}{m_{\tilde t_R}^2 m_{\tilde b_R}^2 m_{\tilde s_R}^2}{\rm Im}[\mta A''^*_{323}] 
\frac{
\Big[ g(x)^2 +\frac{f(x)g(x)}2 \frac{X_t\mta}{m_{\tilde t}^2} + \frac{f(x)^2}{16}\left(\frac{\mta X_t}{m_{\tilde t}^2}\right)^2\Big]
}
{f(x)+8x g(x)\frac{m_t X_t}{ m_{\tilde t}^2}+f(x)\left(\frac{m_t X_t}{ m_{\tilde t}^2}\right)^2
}\nn\\
&+\frac{|\l_{323}|^2g_3^2 \kappa_{\tilde g}}{32\pi^3}
\frac{m_t^2\mta^2}{m_{\tilde t_R}^4\mtg^2}{\rm Im}[\mta \mtg^*] 
\frac{
\Big[ g(x)^2 +\frac{f(x)g(x)}{2} \frac{X_t\mta}{m_{\tilde t}^2} + \frac{f(x)^2}{16}\left(\frac{\mta X_t}{m_{\tilde t}^2}\right)^2\Big]
}
{f(x)+8x g(x)\frac{m_t X_t}{ m_{\tilde t}^2}+f(x)\left(\frac{m_t X_t}{ m_{\tilde t}^2}\right)^2
}\,.\nn
\eeq
Here $x=\frac{m_t}{\mta}$ and the functions $f(x), g(x)$ are defined as follows:
\beq
&f(x)\equiv 1 - 8 x^2 + 8 x^6 - x^8 -   24 x^4 \log x\,,
\\
&g(x)\equiv 1 +9 x^2 -9 x^4 - x^6 +12 (x^2+ x^4) \log x\,.
\eeq
%The leading corrections are $f(x)\simeq 1-8x^2$, $g(x) \simeq 1+9x^2$.
As expected, both the decay rate and the asymmetry parameter decrease with respect to the massless limit, because the available phase space is smaller. For $\mta=500\gev$, the corresponding values are 
\beq
x_0=0.346 ,\qquad f(x_0)=0.42,\qquad g(x_0)=0.24\,. %\Gamma_{exact}(x_0)=0.2\Gamma_0
\eeq

\paragraph{Corrections for effective theory breakdown}
Our results were also computed with the assumption of large squark masses, where the effective field theory approximation is powerful. Because the baryon asymmetry is proportional to $(\mta/\mtq)^3$, the hierarchy between the axino and the squarks cannot be too big; for instance, we took $\mta= \frac59 \mtq=\frac13\mtg$, $\mta= \frac12 \mtq=\frac14\mtg$ and $\mta=\frac12 \mtg=\frac14\mtq$ as concrete benchmark points in eq. \eqref{BP} and in Section \ref{sec:dm}. Thus, it should be verified that corrections to our previous results are small. In the limit of massless final states, the decay rate computed in the full theory is
\beq
&\Gamma_{\rm exact}^{\cancel{\rm EFT}}=\frac{3\kappa_3^2 |\lambda_{323}|^2}{512\pi^3}\frac{m_{\tilde a}^5 m_t^2}{v_{PQ}^2 m_{\tilde t_R}^4}
 \left[1%-8{x g(x)}\frac{m_t X_t}{ m_{\tilde t}^2}
 +\left(\frac{m_t X_t}{ m_{\tilde t}^2}\right)^2
\right]h(y)
\,,\\
&h(y)\equiv 6\frac{6 y^2 - 5 y^4 + 2 (3 - 4 y^2 + y^4) \log(1 - y^2)}{y^8}\,,
%\\&k(y)\equiv \frac{-2 y^2 - (2 - y^2) \log(1 - y^2)}{y^6}\,.
\eeq 
where $y=\frac{\mta}{m_{\tilde t}^2}$ and at small $y$, $h(y)= 1+\frac45y^2+\frac35y^4 +\OO(y^6)$. As expected, the decay rate increases as the axino mass approaches the squark masses. For the asymmetry parameter $\epsilon$, the phase space integral is more complicated and we have to rely on numerical integration.

\begin{figure}[t]
\begin{center}
\subfigure[ ]{
\includegraphics[width=0.43\textwidth]{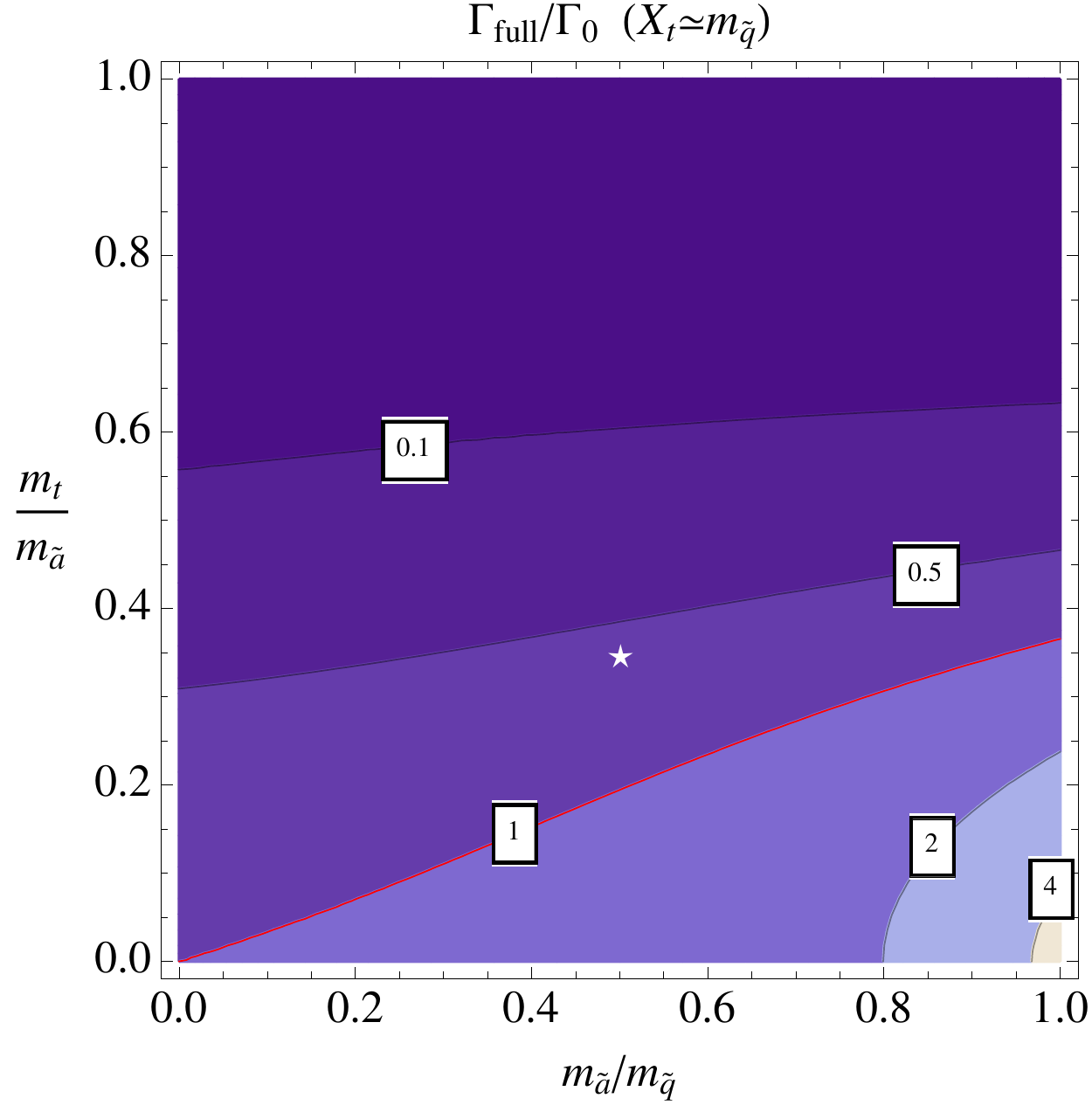}\qquad }
\subfigure[ ]{
\includegraphics[width=0.43\textwidth]{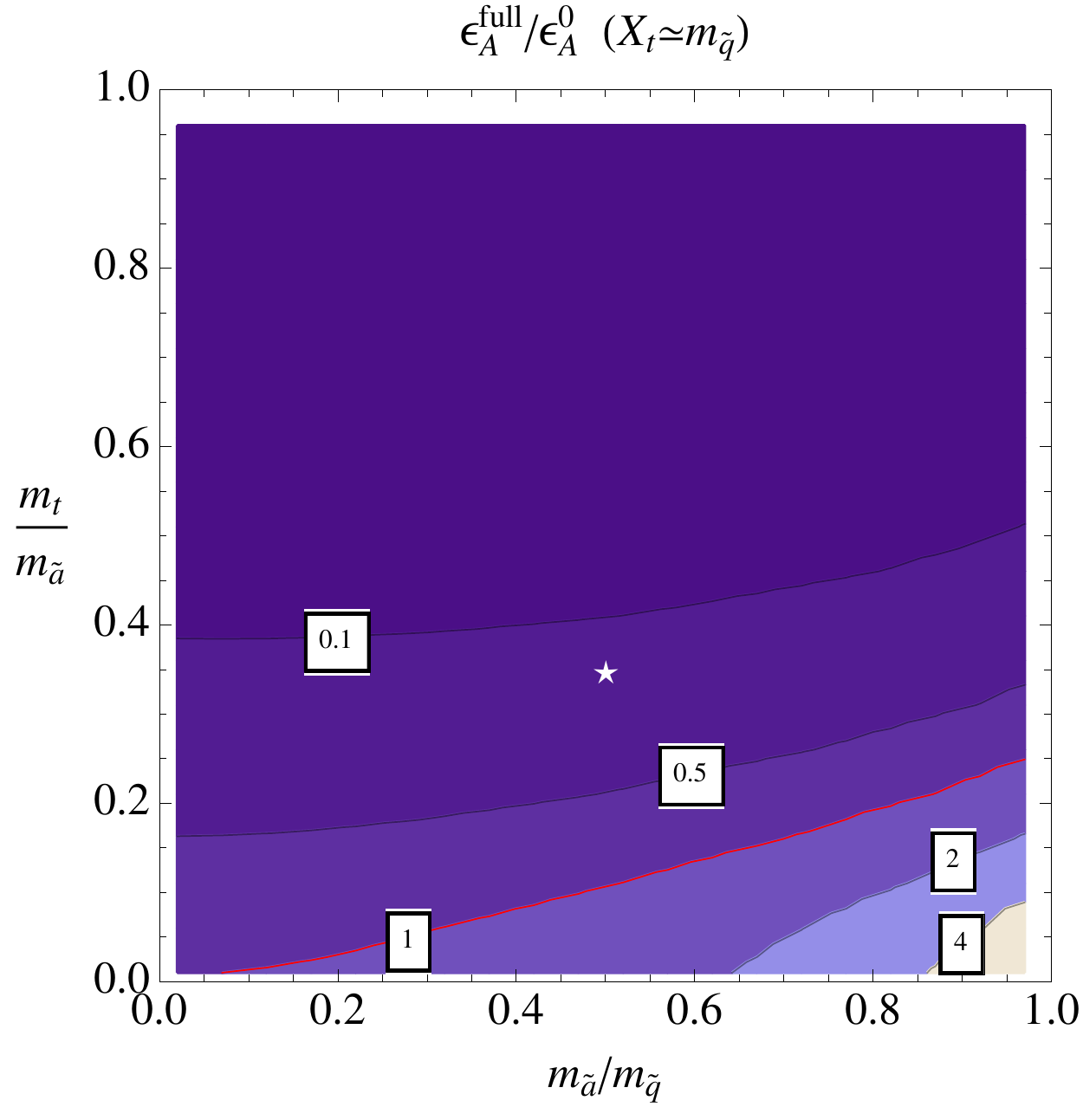}
}
\subfigure[ ]{\includegraphics[width=0.43\textwidth]{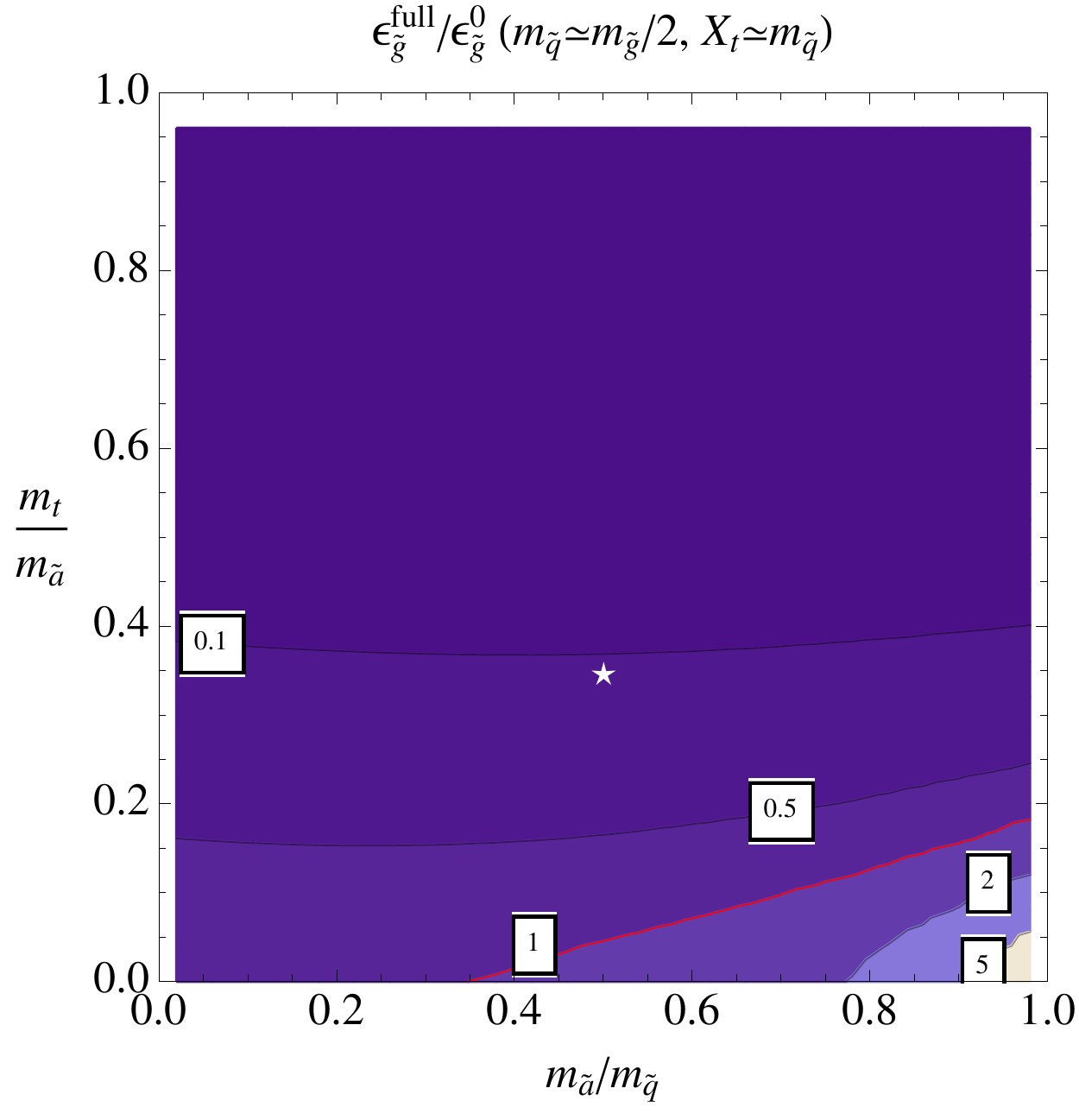}\qquad }
\subfigure[ ]{
\includegraphics[width=0.43\textwidth]{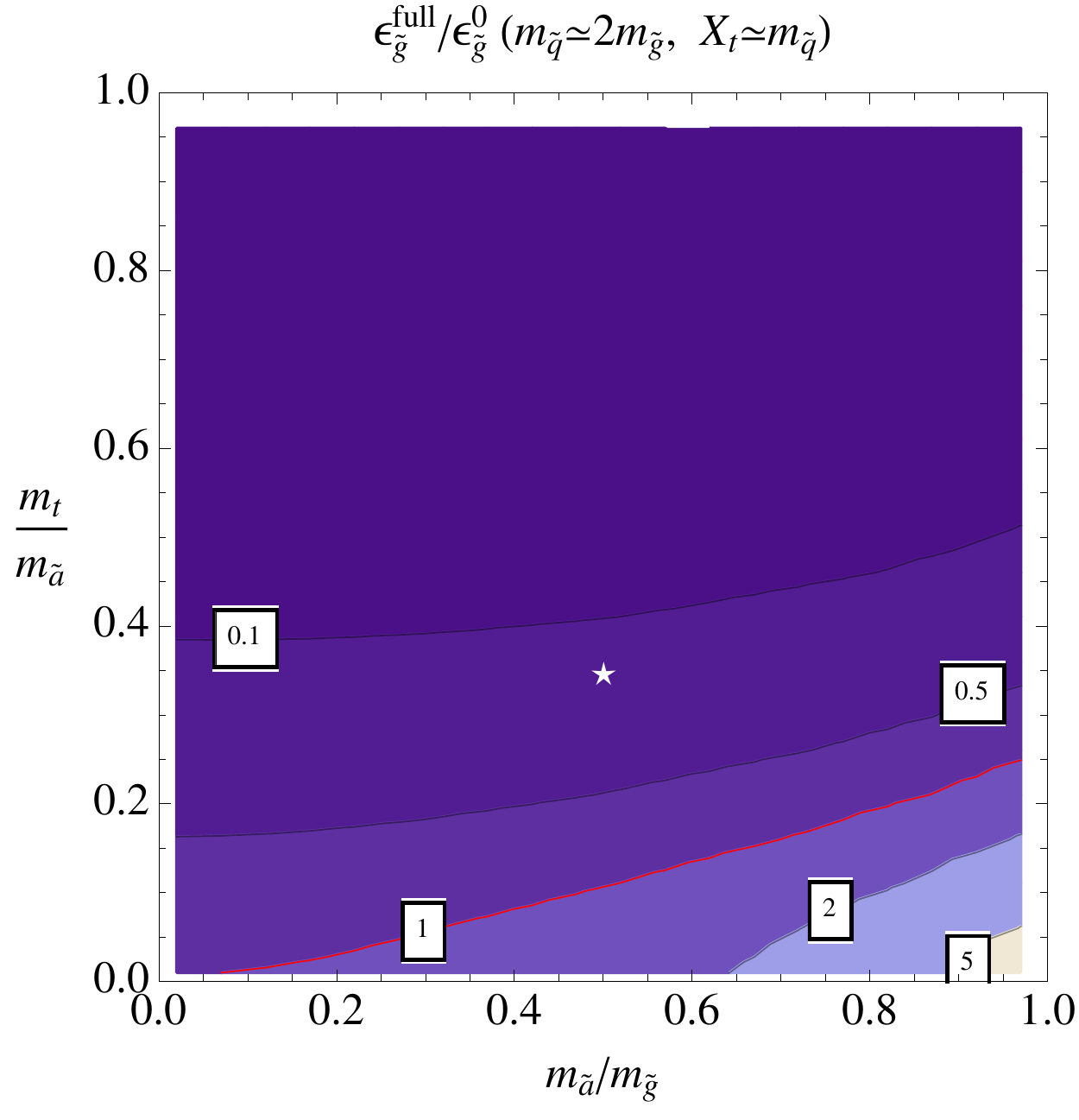}}

\caption{(a) Ratio of $\Gamma_{full}$, the decay rate computed numerically, and $\Gamma_0$, the analytical result of Eq. \eqref{eq:Gaxino}, which was computed in the limit of zero $m_t$ and large squark masses, for $X_t\simeq \mtq$.  (b) Ratio of $\epsilon_A^{full}$ and $\epsilon_A^0$, the contribution to the asymmetry parameter due to $A$-terms. (c,d) Ratio of $\epsilon_{\tilde g}^{full}$ and $\epsilon_{\tilde g}^0$, the contribution to the asymmetry parameter due to gluino exchange. In (c), the gluino has been taken as heavier than the squark, $\mtg=2\mtq$,as at the benchmark point BP2, while in (d) the gluino is lighter, $\mtg=\mtq/2$, as for BP3.
The star markers indicate the benchmark points used in section \ref{sec:dm}. For all figures, the left-right squark mixing has been taken as moderate, $X_t\simeq \mtq$.}
\label{fig:Gamma_numerical}
\end{center}
\end{figure}

\paragraph{Exact numerical results} 
Even in the limit of massless final states, we could not find a simple analytical expression for the asymmetry parameter $\epsilon$ {when taking into account the squark and gluino propagators}. We can  integrate the phase space integrals numerically and check that the simple expressions given in the main part of the article do not introduce  a large error. We show the full numerical results for $\Gamma_{\tilde a}$ and $\epsilon$  in the $\frac{m_t}\mta-\frac\mta \mtq$ plane in Figs. \ref{fig:Gamma_numerical} . The benchmark points of eq. \eqref{BP}, used in Sec. \ref{sec:dm}, are indicated by white star markers.
 We see that $\Gamma$  and $\epsilon$ decrease by about two and six, respectively. This is mainly due to relaxing the approximation $m_t=0$, reducing the phase space available for the decay. The exact numerical results have been used {throughout the paper and} for Fig. \ref{fig:gravDM}.

%\bibliographystyle{JHEP}
%\bibliography{biblio}
%\providecommand{\href}[2]{#2}\begingroup\raggedright

\end{document}